\documentclass[aps,prd,preprintnumbers,amsmath,amssymb,nofootinbib,superscriptaddress,floatfix]{revtex4-2}
\usepackage[latin1]{inputenc}
\usepackage{slashed,xspace,nicefrac}
\usepackage{amsmath,amssymb,amsfonts}
\usepackage{textcomp}
\usepackage{natbib}
\usepackage{enumitem}
\usepackage[section]{placeins}
\usepackage{array,multirow}
\usepackage{threeparttable}
\usepackage{appendix}
\usepackage[table,xcdraw]{xcolor}
\usepackage{color} 
\usepackage{subcaption}
\usepackage{tikz}
\usepackage{makecell}
\usepackage[unicode]{hyperref}
\newcommand{\be}{\begin{equation}}
\newcommand{\ee}{\end{equation}}
\newcommand{\bea}{\begin{eqnarray}}
\newcommand{\eea}{\end{eqnarray}}

\newcommand{\hf}{\nicefrac{1}{2}\xspace}
\newcommand{\nn}{\nonumber\\}

\newcommand{\MAD}{\textsc{MadGraph}\xspace}
\newcommand{\madamc}{\textsc{MadGraph5\_aMC@NLO}\xspace}

\newcommand{\Feyncalc}{\textsc{FeynCalc}\xspace}
\newcommand{\Math}{\textsc{Mathematica}\xspace}
\newcommand{\h}{\ensuremath{\mathcal{H}}}
\newcommand{\hb}{\ensuremath{\overline{\mathcal{H}}}}
\newcommand{\tev}{\mbox{TeV}\xspace}
\newcommand{\gev}{\mbox{GeV}\xspace}
\newcommand{\ifb}{\ensuremath{\mathrm{fb}^{-1}}\xspace}
\usepackage{graphicx,here}
\graphicspath{{figures/}}

\DeclareGraphicsExtensions{.pdf,.png,.jpg}
\graphicspath{{./}{figs/}}
\begin{document}
\begin{flushleft} 
KCL-PH-TH/2023-{\bf 36}
\end{flushleft}

\title{Resummation schemes for high-electric-charge objects  leading to improved experimental mass limits}

\author{Jean Alexandre}
\email{jean.alexandre@kcl.ac.uk}
\affiliation{Theoretical Particle Physics and Cosmology group, Department of Physics, King's College London, London WC2R 2LS, UK}

\author{Nick E.\ Mavromatos}
\email{nikolaos.mavromatos@cern.ch}
\affiliation{Physics Division, School of Applied Mathematical and Physical Sciences, National Technical University of Athens, 15780 Zografou Campus,
Athens, Greece}
\affiliation{Theoretical Particle Physics and Cosmology group, Department of Physics, King's College London, London WC2R 2LS, UK}

\author{Vasiliki A.\ Mitsou}
\email{vasiliki.mitsou@ific.uv.es}
\affiliation{Instituto de F\'isica Corpuscular (IFIC), CSIC -- Universitat de Val\`encia,
C/ Catedr\'atico Jos\'e Beltr\'an 2, 46980 Paterna (Valencia), Spain}
\affiliation{Physics Division, School of Applied Mathematical and Physical Sciences, National Technical University of Athens, 15780 Zografou Campus, Athens, Greece}

\author{Emanuela Musumeci}
\email{emanuela.musumeci@ific.uv.es}
\affiliation{Instituto de F\'isica Corpuscular (IFIC), CSIC -- Universitat de Val\`encia,
C/ Catedr\'atico Jos\'e Beltr\'an 2, 46980 Paterna (Valencia), Spain}

\begin{abstract}
High-Electric-Charge compact Objects (HECOs) appear in several theoretical particle physics models beyond the Standard Model, and are actively searched for in current colliders, such as the Large Hadron Collider at CERN. In such searches, mass bounds of these objects have been placed, using Drell-Yan and photon-fusion processes at tree level so far. However, such mass-bound estimates are not reliable, given that, as a result of the large values of the electric charge of the HECO, perturbative Quantum Electrodynamics calculations break down. In this work, we perform a Dyson-Schwinger resummation scheme (as opposed to lattice strong-coupling approach), which makes the computation of the pertinent HECO-production cross sections reliable, thus allowing us to extract improved mass bounds for such objects from ATLAS and MoEDAL searches. 
\end{abstract}

\maketitle

%%%%%%%%%%%%%%%%%%%%%%%%%%%%%%%%%%%%%%%%%%%%%%%%
%%%%%%%%%%%%%%%%%%%%%%%%%%%%%%%%%%%%%%%%%%%%%%%%
\section{Introduction} 

Several theoretical particle physics models of new physics, beyond the Standard Model (SM), predict sectors with High-Electric-Charge compact solitonic Objects, with finite mass, abbreviated as HECOs. Such objects are highly ionising particles (HIPs), 
and include 
doubly charged massive particles~\cite{dcmp},  
Schwinger dyons~\cite{dyons} (containing both electric and magnetic charges), electrically charged scalars in neutrino-mass models~\cite{scalarneutrino}, Q-balls~\cite{coleman,kusenko}, aggregates of $ud$-quarks~\cite{udaggr}, strange-quark matter~\cite{smatter}, as well as black-hole remnants in models with extra spacetime dimensions~\cite{cbhrem,bhrem}

Throughout this article, the definition of a HECO will be restricted to an object with a {\it sufficiently high value} of its electric charge (in units of the electron charge $e$), such that perturbation theory in the corresponding Quantum Electrodynamics (QED) part of the Lagrangian breaks down. In particular, we shall deal with objects with electric charges $Q \gtrsim 11e$.
Phenomenological aspects of HECOs have been thoroughly studied~\cite{Song}. Experimentally, HECOs are searched for in current colliders, including the Large Hadron Collider (LHC) at CERN by ATLAS~\cite{ATLAS:2015tyu,ATLAS:2019wkg,ATLAS:2023esy} and MoEDAL~\cite{MoEDAL:2021mpi}.\footnote{Detector-stable particles with multiple charge of $\lesssim 10e$ have been searched for at LHC by ATLAS~\cite{ATLAS:2015hau,ATLAS:2018imb,ATLAS:2023zxo} and CMS~\cite{CMS:2013czn}, with very good prospects for the future also by MoEDAL~\cite{Altakach:2022hgn}.} 
In order to place bounds on their masses, based on the experimentally set exclusion HECO-pair production cross-section  limits, the common practice so far in such searches is to use tree-level processes, such as the Drell-Yan (DY) and the photon-fusion (PF) production mechanisms. However, for highly charged HECOs, such an analysis is invalid, due to the associated breakdown of perturbation theory, as a result of the large value of the electric charge involved. 

The situation is similar to that of magnetic monopole searches at colliders~\cite{Mavromatos:2020gwk}, in which also perturbation theory fails miserably, due to the large value of the magnetic charge of the monopole, which is equivalent to charges of size $68.5e$ or more. We mention for completeness at this point that, in the case of production of magnetic monopole--antimonopole pairs from, say, proton--proton ($pp$) collisions at the LHC, one may recover perturbation theory if one considers the so-called $\beta$-dependent effective monopole coupling, required by electromagnetic duality~\cite{milton1,milton2,baines}, where $\beta$ is an appropriate Lorentz invariant ``velocity'' (expressed in units of the speed of light in vacuo $c$). Perturbation theory is recovered in the case of slowly moving monopoles (i.e.\ $\beta \ll 1$), allowing for reliable mass upper bounds for the monopole to be obtained in this case from experimental upper limits on cross sections~\cite{baines}.

In the case of HECOs, however, such duality arguments are absent, and, thus, one does not have $\beta$-dependent electric charges. One needs therefore to resort to non-perturbative studies, e.g.\ strongly coupled QED on the lattice, or continuous Dyson-Schwinger (DS) resummation schemes.
The latter approach provides exact relations between the fully dressed propagators and vertices of a gauge theory, 
in the form of integral equations. After choosing an ansatz for the functional dependence of the effective action (consistently with Ward identities),
these coupled equations provide a partial resummation of the effective theory, which allows the derivation of non-perturbative effects. 
An example is the dynamical mass generation for fermions, in the process of magnetic catalysis~\cite{Gusynin:1999pq}: this dynamical mass is not analytical in the coupling constant, and can be found only from the resummation of an infinite number of Feynman graphs. 
Alternatively, the use of DS equations also leads to a description of the highly non-trivial infrared dynamics in Quantum Chromodynamics,
and the presence of a gauge-invariant pole in the gluon propagator~\cite{Aguilar:2015bud}, which leads to confinement. 
This effect involves a polarisation tensor which is not analytical in the gluon momentum, and such a result cannot be found with a perturbative approach.
We mention for completeness that such DS resummations have been previously applied also to the magnetic monopole case~\cite{AM}. The important difference though of the HECO case from that of the magnetic monopole lies on the fact that in the former case one knows the relevant part of the effective QED Lagrangian, which describes the electromagnetic interactions of the HECO. 
Par contrast, at least currently, the effective Lagrangian of the magnetic charge is largely unknown.

In Ref.~\cite{AM}, we have conjectured that the quantum interactions of a magnetic monopole with ordinary photons, which enter the aforementioned DY and PF processes (adapted to the case of monopole-pair production) are described by an appropriately constrained $U(1)_\text{weak} \times  U(1)_\text{strong}$ gauge theory, where $U(1)_\text{weak}$ denotes the standard gauge theory of ordinary electromagnetism, that is QED, whilst $U(1)_\text{strong}$ is a strongly coupled Abelian gauge theory which is associated with the magnetic charge $g_m$ of the monopole. For this latter part of the theory,  a distinct non-perturbative DS resummation scheme has been employed, which leads, among other properties, also to the appearance of an effective form factor in the interaction cross section of a magnetic monopole with ordinary matter~\cite{AM}. This has been interpreted as the aforementioned effective $\beta$-Lorentz-invariant velocity factor that 
was conjectured to dress the effective magnetic charge, $\beta g_m$
when a monopole is considered in interaction with ordinary matter, as a consequence of equivalence under electric--magnetic duality of the monopole--matter interaction to the Rutherford scattering between two matter charges~\cite{milton1,milton2,baines}. 

In this article we shall extend the DS resummation method of Ref.~\cite{AM} to the case of highly charged HECOs. The extension contains differences from the magnetic monopole case. Among others, in the HECO resummation case, in contrast to the magnetic-monopole case, one recovers standard perturbative QED results when the HECO charge is lower than $\simeq 11e$.
For computational ease, we shall restrict ourselves to the case where the HECO is a spin-\hf fermion. Extension to the scalar case is straightforward, although algebraically more complex, as it involves more graphs. It does not, however, present any conceptual difficulties. Finally, we also mention, for completeness, that the spin-one case presents similar issues with unitarity as the effective field theory of the charge $W$-bosons of the SM, in which case unitarity is restored only for a specific value $\kappa=1$ of the magnetic dipole moment.  

In the case of spin-\hf HECOs we shall show that the resummation leads to an effective QED-like Lagrangian for the highly charged HECOs, in which the effective charge and mass of the object are given by appropriate resummed quantities involving 
the wave-function renormalisation of the corresponding dressed HECO-fermion propagation. Our analysis will employ a specific renormalisation scheme, and certain approximations similar in nature to the DS analysis of the magnetic-monopole case~\cite{AM}. It should not, therefore, be considered as the most general study. Nonetheless, as we shall demonstrate, it leads to a significant improvement of the mass bounds as compared to the previously set mass bounds using only the tree-level DY and PF HECO pair-production processes. Our analysis will also include the coupling of the HECO to the $Z^0$-boson of the SM~\cite{Song}. For brevity and concreteness we shall only consider here non-chiral couplings of the spin-\hf HECO to photons and $Z^0$-bosons. Resummation of more general (chiral) couplings of HECOs to other gauge bosons that can appear in extensions of the SM is straightforward, albeit more technically involved,  and does not present any conceptual difficulties. 

The structure of the article is as follows. In  Section~\ref{sec:DS}
we develop the DS resummation scheme for the aforementioned simplified models of spin-\hf HECO, and arrive at the effective Lagrangian for the HECO, including interactions of the HECO with both photons and $Z^0$ bosons. We give the appropriate non-perturbative solution of the DS equations, which involves appropriate running couplings and masses, depending on the energy scale $k$. These parameters are then substituted in appropriately dressed tree-level DY and PF processes in Section~\ref{sec:hecorel}.
In Section~\ref{sec:feyn}, we implement the resummation effects in existing Feynman-rules algorithms, used subsequently to estimate production cross sections at the LHC using the simulation tool \MAD~\cite{madgraph}. 
The resulting cross sections are confronted with the experimental searches results in Section~\ref{sec:limits}, leading to the extraction of stronger mass bounds for HECOs. Conclusions and outlook are discussed in Section~\ref{sec:concl}. Finally, the Feynman rules incorporated in \MAD and the validation of the Monte Carlo simulation against  analytical calculations are presented in Appendices~\ref{feynman_appendix} and~\ref{validation_appendix}, respectively.

%%%%%%%%%%%%%%%%%%%%%%%%%%%%%%%%%%%%%%%%%%%%%%%%
%%%%%%%%%%%%%%%%%%%%%%%%%%%%%%%%%%%%%%%%%%%%%%%%
\section{Strong coupling solution of Dyson-Schwinger equations}\label{sec:DS} 

In this section we present our DS resummation scheme which itself employs  specific approximations that we state explicitly. It is therefore not the more general solution, but it suffices to provide already significant improvement in the derivation of mass bounds obtained from comparing the results of our effective Lagrangian to experimentally acquired upper limits on HECO-production cross sections at current colliders, as compared to previous tree-level processes of Drell-Yan or photon-fusion type used in the interpretation of data so far. 
%In what follows we restrict ourselves to non-chiral coupling of spin-\hf HECO to photons and $Z^0$-bosons. 
We derive the  DS resummation for the effective HECO mass and coupling,  which run with the renormalisation group momentum scale, and compute those at the ultraviolet (UV) fixed-point solution, which is used to construct our effective Lagrangian to be used in the theoretical cross section computations, which will take place in the following Section~\ref{sec:hecorel}.

%%%%%%%%%%%%%%%%%%%%%%%%%%%%%%%%%%%%%%%%%%%%%%%%
\subsection{Coupled quantum corrections}

We consider the electromagnetic interactions of a HECO, 
assumed for concreteness to be a spin-\hf Dirac fermion, $\psi (x)$, 
which couples to a massless photon $A_\mu (x)$ with a (bare) charge
\begin{align}\label{hecoch}
g= n e\,, \quad n \in \mathbb Z^+ \, \,\,({\rm a~positive~ integer}),
\end{align}
with $e$ the electron charge. The bare Lagrangian is given by the standard QED Lagrangian:
\bea\label{qedlag}
\mathcal L_\text{bare} = -\frac{1}{4} F_{\mu\nu}\, F^{\mu\nu} + \frac{\lambda}{2} (\partial_\mu A^\mu)^2
+ \overline \psi \Big(i\slashed \partial + g\slashed A  - m \Big) \, \psi\,, 
\eea
with $m$ the bare HECO mass and $\lambda \in \mathbb R$ a gauge fixing parameter.
The dressed propagators and vertex are assumed to have the form
\bea\label{ansatz}
G&=&i\frac{\mathcal Z\slashed p+M}{\mathcal Z^2p^2-M^2},\nonumber \\
\Delta_{\mu\nu}&=&\frac{-i}{(1+\omega)q^2}\left(\eta_{\mu\nu}+\frac{1+\omega-\lambda}{\lambda}\frac{q_\mu q_\nu}{q^2}\right),\nn
\Gamma_\mu&=&g\, \mathcal Z\, \gamma_\mu~,
\eea
where the quantum corrections $\mathcal Z,\omega,M$ are momentum-independent.\footnote{In this article we work with the Minkowski spacetime metric $\eta^{\mu\nu}$ with signature $(+,-,-,-)$.} 

We then consider the set of coupled Dyson-Schwinger equations for the photon and the fermion self-energies, 
in dimension $4-\epsilon$, for which the ansatz (\ref{ansatz}) leads to \cite{AM}
\bea\label{DS}
\mathcal Z&=&1+\frac{g^2}{8\pi^2\lambda}\frac{1}{\epsilon}\left(\frac{\mathcal Zk}{M}\right)^\epsilon+\mbox{finite},\nonumber \\
\omega&=&\frac{g^2}{6\pi^2\, \mathcal Z}\frac{1}{\epsilon}\left(\frac{\mathcal Z\,k}{M}\right)^\epsilon+\mbox{finite},\nn
1-\frac{m}{M}&=&\frac{g^2}{8\pi^2\lambda \, \mathcal Z}\frac{1+3\lambda+\omega}{1+\omega}
\frac{1}{\epsilon}\left(\frac{\mathcal Z\,k}{M}\right)^\epsilon+\mbox{finite},
\eea
where $m$ is the fermion bare mass and $k$ is an energy scale introduced by dimensional regularisation.
Our aim is to solve the full set of equations, without assuming a perturbative regime $g^2 \ll 1$, 
and therefore allowing for large corrections $(\mathcal Z-1,~\omega,~M-m)$.

In what follows, {\it we assume that $g$ and $m$ are independent of the scale $k$}, and we 
take a derivative of Eqs.~\eqref{DS} with respect to $k$ in order to eliminate $\epsilon$. 
We then integrate back with respect to $k$, which has the effect to swap the $\epsilon$-dependence with constants of integration. 
We choose boundary conditions such that 
\begin{align}\label{bc}
(\mathcal Z,\,\omega,\, M)=(1,\,0,\, m),\quad {\rm when}\quad  k=m\,, 
\end{align}
which lead to
\bea\label{ZomegaM}
\mathcal Z&=&1+\frac{g^2}{8\pi^2\lambda}\ln\left(\frac{\mathcal Z\,k}{M}\right), \nonumber \\
\mathcal Z\omega&=&\frac{g^2}{6\pi^2}\ln\left(\frac{\mathcal Z\,k}{M}\right),\nn
\mathcal Z\left(1-\frac{m}{M}\right)&=&\frac{g^2}{8\pi^2\lambda}\frac{1+3\lambda+\omega}{1+\omega}\ln\left(\frac{\mathcal Z\,k}{M}\right)~.
\eea
We now note that the Eqs.~\eqref{ZomegaM} recover the usual one-loop results for $g^2 \ll 1$:
\bea\label{logevol}
\mathcal Z&=&1+\frac{g^2}{8\pi^2\lambda}\ln\left(\frac{k}{m}\right)+{\cal O}(g^4),\\
\omega&=&\frac{g^2}{6\pi^2}\ln\left(\frac{k}{m}\right)+{\cal O}(g^4),\nn
M&=&m+\frac{g^2m}{8\pi^2\lambda}(1+3\lambda)\ln\left(\frac{k}{m}\right)+{\cal O}(mg^4)~.\nonumber
\eea

%%%%%%%%%%%%%%%%%%%%%%%%%%%%%%%%%%%%%%%%%%%%%%%%
\subsection{Fixed point solution}

In what follows we focus on the Feynman gauge 
\bea\label{feyngauge}
\lambda=1\,, 
\eea
which corresponds to a ``physical'' gauge in the framework of scattering 
processes. This can be justified perturbatively by the resummation of a class of Feynman graphs, where gauge dependence cancels out to lead to 
a result obtained with the Feynman gauge (see ``pinched technique'' \cite{pinched,pinched2}). 

\begin{figure}[ht]
    \includegraphics[width=0.6\linewidth]{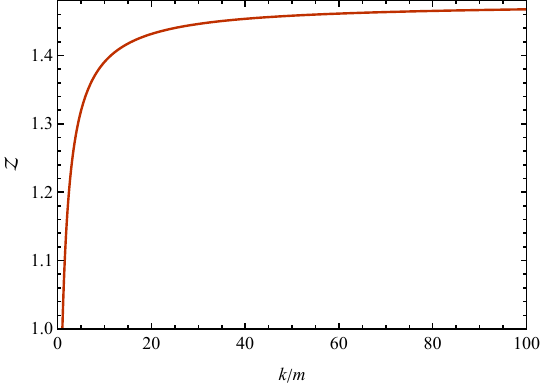}
    \caption{The wave-function renormalisation $\mathcal Z$ as a function of $k/m$ for $g^2=8\pi^2$. $\mathcal Z$ asymptotically goes to $\mathcal Z^\star\equiv \mathcal Z_+$ when $k\to\infty$.}
    \label{Zfig}
\end{figure}

Because of the resummation provided by the DS approach, the Eqs.~\eqref{ZomegaM} are also valid for strong coupling $g$,
in which case one should keep $\mathcal Z$ and $M$ in the argument of the logarithm in Eqs.~\eqref{ZomegaM}, leading to a modification
of the naive logarithmic evolution in Eqs.~\eqref{logevol}. We show here that one can find an UV fixed-point solution of Eqs.~\eqref{ZomegaM},
where the running mass $M/\mathcal Z$ goes to infinity, but in such a way that 
\be
\lim_{k\to\infty}\frac{k\,\mathcal Z}{M}=~\mbox{finite}~,
\ee
and where $\mathcal Z$ and $\omega$ are finite, too.
The Eqs.~\eqref{ZomegaM} can be written in the form
\bea\label{ZomegaMbis}
\mathcal Z&=&1+\frac{g^2}{8\pi^2}\ln\left(\frac{\mathcal Z\,k}{M}\right),\nonumber \\
\omega&=&\frac{4}{3}\left(1-\frac{1}{\mathcal Z}\right),\nn
1-\frac{m}{M}&=&4~\frac{4\mathcal Z-1}{7\mathcal Z-4}\left(1-\frac{1}{\mathcal Z}\right)~,
\eea
and one can express $k$ in terms of $\mathcal Z$
\be\label{k}
\frac{k}{m}=\frac{7\mathcal Z-4}{9(\mathcal Z_+- \mathcal Z)(\mathcal Z-\mathcal Z_-)}~\exp\left(\frac{8\pi^2}{g^2}(\mathcal Z-1)\right)~,
\ee
where 
\be
\mathcal Z_\pm=\frac{8}{9}\left(1\pm\frac{\sqrt7}{4}\right)\simeq
\begin{cases}
1.477=\mathcal Z^\star \\
0.301
\end{cases}~.
\ee
One can see that the expression (\ref{k}) for $k$ is positive for $\mathcal Z_-< \mathcal Z< \mathcal Z_+$, and that $k\to\infty$ when $\mathcal Z\to \mathcal Z_\pm$.
Hence a consistent solution, in agreement with the unitarity requirement for the fermion $\mathcal Z > 1$ sector, which, 
on account of Eq.~\eqref{ZomegaMbis} also implies $\omega > 0$, is given by:
\begin{enumerate}[label={(\roman*)}]
    \item boundary condition at $k=m$, with $\mathcal Z=1$;
    \item $\mathcal Z\to \mathcal Z^\star$ when $k\to\infty$, which is depicted in Figure~\ref{Zfig}.
\end{enumerate}
In the limit $k\to\infty$, we also have 
\be\label{omstasrdef}
\omega\to\omega^\star=\frac{4}{3}\left(1-\frac{1}{\mathcal Z^\star}\right)\simeq0.431~,
\ee
and for the dimensionless mass parameter
\be\label{mhgonly}
\tilde M\equiv\frac{M}{k}\to \mathcal Z^\star\exp\left(-\frac{8\pi^2}{g^2}(\mathcal Z^\star-1)\right)~,
\ee
corresponding to the UV fixed point 
\be\label{fixedpoint}
\lim_{k\to\infty}(\mathcal Z,\omega,\tilde M)=(\mathcal Z^\star,\omega^\star,\tilde M^\star)~.
\ee
This novel UV fixed point, which to our knowledge, has not been studied in QED before, refers to large masses, since it involves the limiting situation $M \to \infty$ as $k \to \infty$, such that $M/k$ remains finite.

We note that the property $\mathcal Z\ge1$, consistent with unitarity, allows in principle the HECO to be a fundamental particle. This is to be contrasted with the work of Ref.~\cite{AM}, where the same DS equations are solved in the regime $\mathcal Z \le 1$, which is possible for a composite state only.

%%%%%%%%%%%%%%%%%%%%%%%%%%%%%%%%%%%%%%%%%%%%%%%%
\subsection{Running effective coupling}\label{sec:rc}

The Lagrangian based on the running effective parameters is
\bea\label{renlag}
{\cal L}_\text{eff} =\frac{1}{2}A_\mu\Big((1+\omega)\eta^{\mu\nu}\Box-\omega\partial^\mu\partial^\nu\Big)A_\nu 
+\overline\psi\Big(\mathcal Zi\slashed\partial+\mathcal Zg\slashed{A}-M\Big)\psi~.\eea
We then perform the simultaneous rescalings 
\be\label{resc}
A_\mu\to A_\mu/\sqrt{1+\omega}~~~~\mbox{and}~~~~\psi\to\psi/\sqrt{\mathcal Z}~, 
\ee
to obtain the canonically normalised Lagrangian
\bea\label{efflag}
{\cal L}_\text{eff}\to\frac{1}{2}A_\mu\left(\eta^{\mu\nu}\Box -\frac{\omega}{1+\omega}\partial^\mu\partial^\nu\right)A_\nu +\overline\psi\left( i\slashed\partial+\frac{g}{\sqrt{1+\omega}}\slashed A-\frac{M}{\mathcal Z}\right)\psi~.
\eea
Thus we observe that, for a fixed $k$, the effective  Lagrangian \eqref{efflag},  corresponding to the above-described DS resummation, is nothing other than a QED Lagrangian, but with a ``shifted'' covariant gauge parameter
\bea\label{shiftcg}
\lambda_\text{eff} = \frac{1}{1 + \omega}= 1 - \frac{\omega}{1+\omega} 
\eea
and a running HECO-fermion mass, ${\cal M}(k)$, and effective running coupling, $g(k)$, given, respectively, by
\be
{\cal M}(k)=\frac{M(k)}{\mathcal Z(k)}~, \quad
\alpha(k)=\frac{g^2/(4\pi)}{1+\omega(k)}~,
\ee
where, as usual, we gave explicitly the fine structure constant combination for the coupling.
In view of Eq.~\eqref{shiftcg}, the reader should expect that physical quantities pertaining to processes, such as cross sections or decay rates, should not depend on the use of the one-loop-resummed or bare photon propagator. This is in fact confirmed explicitly, as we shall discuss in Section~\ref{sec:feyn}.

We also note that in the usual approach one fixes $\alpha\equiv\alpha_0$ at a given scale $k_0$, and then one infers the evolution of the bare coupling $g_b(k)$
with the scale $k$. In the perturbative case where $\omega(k)\ll1$, this procedure leads to
\be
\frac{g^2_b(k)}{4\pi}=\big(1+\omega(k)\big)\alpha_0\simeq\frac{\alpha_0}{1-\omega(k)}~,
\ee
which reproduces the Landau pole with the perturbative expression for $\omega(k)$ given in Eqs.~\eqref{logevol}.
The two approaches differ by the boundary condition which is chosen, and by the non-perturbative resummation.
In the present case, where we start with the fixed bare coupling $g$, 
non-perturbative quantum fluctuations lead to the finite dressed coupling $\alpha(k)$, for all values of $k$.

The non-perturbative solution arising from the DS resummation therefore 
avoids a divergence in the form of the Landau pole. This result is similar to an alternative non-perturbative approach, developed in Ref.~\cite{APS}, 
based on a differential functional equation, which describes the evolution of the dressed system with the bare mass.
In that work, the Landau pole is also avoided, as a consequence of a non-trivial resummation of quantum corrections.

%%%%%%%%%%%%%%%%%%%%%%%%%%%%%%%%%%%%%%%%%%%%%%%%
%%%%%%%%%%%%%%%%%%%%%%%%%%%%%%%%%%%%%%%%%%%%%%%%
\section{Relevance to Objects with High Electric Charge} \label{sec:hecorel}

In this section we implement the previous results to the construction of the ``running'' mass of a HECO, denoted by \h, as a function of  the renormalisation group momentum scale, $k$. Then we use the results to construct the appropriate Feynman rules that describe the effective (resummed) non-chiral interactions of the spin-\hf HECO, first with the photon and then with a linear combination of photons and $Z^0$-bosons in the SM sector. The scale $k$ is identified with the UV cutoff $\Lambda$ that here plays the r\^ole of the energy scale at which new physics appears. In the case of collider searches, this can be taken to be the centre-of-mass energy, $\sqrt{s}$, at which the current high-energy collider, LHC, operates, that is $\mathcal{O}(10~\tev)$.

%%%%%%%%%%%%%%%%%%%%%%%%%%%%%%%%%%%%%%%%%%%%%%%%
\subsection{Running mass}\label{sec:rm}

From the previous results, we obtain the flow equation for the  mass scale 
\be
{\cal M}(k)=k\exp\left(-\frac{2\pi}{\alpha(k)}(\mathcal Z(k)-1)\right)~,
\ee
which runs from ${\cal M}(m)=m$, acquired for $k=m$, to reach asymptotically the value
\be
\lim_{k \to \Lambda} {\cal M}(k ) \equiv {\cal M}(\Lambda)=\Lambda\exp\left(-\frac{2\pi}{\alpha^\star}(\mathcal Z^\star-1)\right)~
\label{Lambda},
\ee
where $\Lambda\gg m$ is a natural UV cutoff in the context of these studies.
The boundary condition \eqref{bc} 
implies that the flow in $k$ provides an interpolation between the bare (infrared) and the renormalised mass of the HECO particle, 
corresponding to the UV fixed point.
If the HECO charge coincides with that of the electron one obtains a perturbative QED situation in the infrared, 
but in that case the HECO does not necessarily correspond to the electron.

\begin{figure}[ht]
    \includegraphics[width=0.6\linewidth]{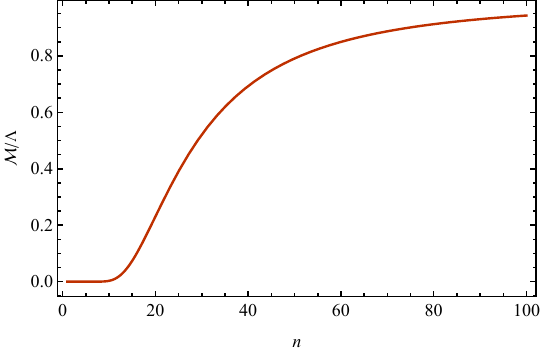}
    \caption{${\cal M}(\Lambda)/\Lambda$ as a function of $n$, as given by Eq.~\protect\eqref{MLambda}. The ratio asymptotically goes to unity when $n\to\infty$.}\label{Mfig}
\end{figure}

If the HECO charge is given by Eq.~\eqref{hecoch}, then we have
\bea\label{effcoupl}
\frac{2\pi}{\alpha^\star}(\mathcal Z^\star-1)
\simeq\frac{2\pi\times137\times(1+0.431)\times(1.477-1)}{n^2}~,
\eea
and therefore 
\be\label{MLambda}
\frac{{\cal M}(\Lambda)}{\Lambda}\simeq\exp\left(-587/n^2\right)~,
\ee
which is drawn in Figure~\ref{Mfig}. Hence, the ``resummed'' mass ${\cal M}(\Lambda)$ is defined by the cutoff scale $\Lambda$, which in turn is a free parameter loosely connected to the new-physics scale within the framework of an effective field theory for the HECOs. In the context of the production cross-section calculation, $\Lambda$ spans the parton collision energy range, that is $\sim 1$--$10~\tev$ in the LHC case.

\begin{figure}[ht]
\centering
\begin{subfigure}[b]{0.27\textwidth}
         \centering
         \includegraphics[width=\textwidth]{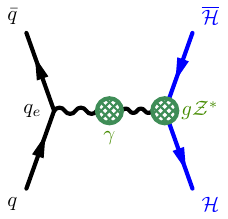}
         \caption{Drell-Yan, $\gamma$ exchange}
         \label{fig:dy-gamma}
     \end{subfigure}
     \hspace{0.02\textwidth}
     \begin{subfigure}[b]{0.27\textwidth}
         \centering
         \includegraphics[width=\textwidth]{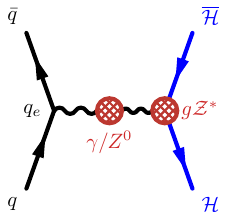}
         \caption{Drell-Yan, $\gamma/Z^0$ exchange}
         \label{fig:dy-gamma-z}
     \end{subfigure}
     \hspace{0.02\textwidth}
     \begin{subfigure}[b]{0.31\textwidth}
         \centering
         \includegraphics[width=\textwidth]{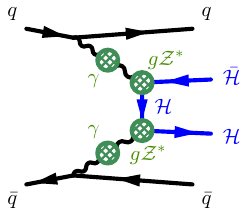}
         \caption{Photon fusion}
         \label{fig:photon-fusion}
     \end{subfigure}
    \caption{Considered processes for the production of spin-\hf HECO-anti-HECO (\h\hb) pairs at hadron colliders: (a) Drell-Yan with only $\gamma$ exchange; (b) Drell-Yan also including $Z^0$-boson exchange; and (c) photon-fusion processes for the production of spin-\hf HECO-anti-HECO (\h\hb) pairs at hadron colliders. The green (red) blobs denote the resummed photon (combined $\gamma/Z^0$-boson) propagators and \h--\hb--$\gamma$ (\h--\hb--$\gamma/Z^0$) vertices, implied by Eq.~\protect\eqref{feynmanrules}. }
    \label{fig:process}
\end{figure}

At this UV fixed point, one can calculate the HECO production cross sections at colliders. To this end, we use Feynman rules obtained from the effective Lagrangian \eqref{efflag}, in which case Eqs.~\eqref{omstasrdef}, \eqref{shiftcg}
\eqref{effcoupl} and \eqref{MLambda} are valid. That is, we use the HECO-fermion and photon propagator, as well as the photon-HECO-fermion vertex, as 
\bea\label{feynmanrules}
G^\text{eff}&=&i\frac{\slashed p+\mathcal M(\Lambda)}{p^2-\mathcal M(\Lambda)^2}\nonumber\\
\Delta^\text{eff}_{\mu\nu}&=&\frac{-i}{q^2}\left(\eta_{\mu\nu}+ \omega^\star\, \frac{q_\mu q_\nu}{q^2}\right)\nn
\Gamma^\text{eff}_\mu&=&g \, \mathcal Z^\star\gamma_\mu~,
\eea
respectively. To avoid overcounting of the resummed loops, we need to restrict ourselves to tree-level Drell-Yan or photon-fusion processes, of the type discussed in Refs.~\cite{Song,MoEDAL:2021mpi,ATLAS:2023esy} and depicted in Figure~\ref{fig:process}, 
but with the important replacement of the standard tree-level Feynman rules by the above, effective ones \eqref{feynmanrules}. As we can see from the above discussion, the order of magnitude of the corresponding graphs obtained by the DS resummation described above, is the same with the ones calculated using the corresponding naive tree-level processes as in Refs.~\cite{Song,MoEDAL:2021mpi,ATLAS:2023esy}. 

A discussion is in order at this point on the modified $\h-\hb-\gamma$ vertex equation in \eqref{feynmanrules}. Naively, from the effective Lagrangian \eqref{efflag}, at the UV fixed point,  one would expect to use the tree-level vertex rule $\Gamma_\mu = \frac{g}{\sqrt{1 + \omega^\star}} \, \gamma_\mu$. 
 However, the Lagrangian \eqref{efflag} is {\it gauge fixed} due to the non-trivial $\omega^\star$-dependent longitudinal terms ($\propto \partial^\mu\partial^\nu$) of the gauge sector. Hence, at the level of the renormalised Lagrangian \eqref{renlag}, before the rescalings \eqref{resc}, 
 the standard Ward identity, stemming from gauge invariance in conventional QED, which necessitates the equality of the wave-function rernormalizations for the HECOs ($\mathcal Z$) and the vertex ($\mathcal Z_V$), $\mathcal Z = \mathcal Z_V$, is no longer applicable in the resummed strongly-coupled QED case at hand. This implies that, following the standard treatment, one can define a \emph{renormalised} HECO coupling $g_R$ (at the UV fixed point) by the rescaling:
\begin{align}\label{hecoRcoupl}\, 
g \to g_R\, \sqrt{(1 + \omega^\star)}\, \frac{{\mathcal Z}^\star}{{\mathcal Z_V}^\star}\,,
\end{align}
which, upon the appropriate field rescalings by the pertinent wave-function renormalisations in the UV-fixed-point effective Lagrangian, so as to have fields with canonical kinetic terms, leads to 
$g \to \frac{1}{\mathcal Z_V^\star} g_R$. We expect naturally ${\mathcal Z^\star_V}^{-1} \sim \mathcal Z^\star \sim 1$, which leads to the generalised vertex rule of \eqref{feynmanrules} (omitting, from now on, the suffix $R$ for notational brevity). A complete treatment would require analysing also the resummed vertex DS equation, together with the resummed photon and HECO-self-energy DS equations, This would lead to a self consistent DS system of equations, which could be solved for the three wave-function renormalisations at the UV fixed point. In the current work we omit such a more complete analysis, presenting instead, only the above phenomenological argument.   
The reader should also notice, for completion, that the photon wave-function renormalisation factor $\sqrt{1+\omega^\star}$ can be absorbed in the definition of the electron charge, in units of which the (renormalised) HECO charge is expressed.
This can be seen by including the electron and quark sector couplings with  the electromagnetic (photon) field, and rescaling the latter by its  wave-function renormalisation at the UV fixed point, so as to guarantee the canonically-normalised Maxwell term in \eqref{efflag}.

%%%%%%%%%%%%%%%%%%%%%%%%%%%%%%%%%%%%%%%%%%%%%%%%
\subsection{Inclusion of the $Z^0$ boson}\label{sec:zb}

Unlike the case of magnetic monopoles, which are assumed to interact only with photons, i.e.\ with the electromagnetic neutral current, HECOs could also interact via the weak neutral current, that is they are characterised by non-trivial couplings with the SM $Z^0$-boson. 

In standard perturbative treatments~\cite{Song}, the coupling of the HECO to $Z^0$-boson requires some assumptions to be made, concerning the nature of the transformation of HECOs under the weak-isospin gauge group $SU(2)$ of the SM. 
In principle, one may assume a coupling of the HECO with the entire $SU(2) \times U(1)_Y$ of the SM, which can be implemented via the appropriate covariant derivative acting on our spin-\hf HECO field, in a Dirac-like Lagrangian term. 
However, because the gauge group $SU(2)$ is chiral, acting only on the left-handed HECO fermions, there will be, in general, asymmetric couplings between left $(L)$ and right $(R)$ sectors. The corresponding \h--\hb--$Z^0$  interaction Lagrangian can thus be expresses as~\cite{Song}:
\begin{align}\label{chiralHECO}
\mathcal L_\text{int} &= - \frac{e}{\sin(2\,\theta_W)} \hb \, \gamma^\mu Z^0_\mu \, \Big(c_L \, \frac{1}{2} (1-\gamma^5) + c_R \, \frac{1}{2} (1 + \gamma^5)\Big)\, \h \nonumber \\ 
&= - \frac{e}{\sin(2\,\theta_W)}\,\hb\, \gamma^\mu Z^0_\mu \, \Big[\frac{1}{2} (c_L + c_R)  -  \frac{1}{2}(c_L-  c_R)\, \gamma^5\,\Big]\, \h\,,
\nonumber \\
c_{L} &= t^3 - |n|\, \sin^2\theta_W\,, \quad c_R = - |n|\, \sin^2\theta_W\,,
\end{align}
where $|n|$ is the HECO charge multiplicity, $\theta_W$ is the weak mixing angle, and $t^3$ is the value of the third component  of the weak isospin gauge group $SU(2)$. 
The usual assumption is that a HECO is a singlet under such $SU(2)$ transformations, so a HECO fermion is an eigenstate of $t^3$ with zero eigenvalue $t^3=0$. In this case, Eq.~\eqref{chiralHECO} is simplified to a purely vector interaction:
\begin{align}\label{nonchiralHECO}
\mathcal L_\text{int} = \frac{1}{2}\, e\, |n|\tan\theta_W\,\hb \, \gamma^\mu Z^0_\mu  \, \h\,.
\end{align}
We estimate below the contribution of the weak boson $Z^0$ to the non-perturbative process in view of Eq.~\eqref{nonchiralHECO}.
The DS equation for the fermion self-energy has the structure of a one-loop correction, in which, however, the propagators and one of the vertices are dressed, which is the essence of the resummation of quantum corrections.
Hence the $Z^0$ contribution is additive, and also has the structure of a one-loop graph, but with dressed parameters in the loop.
The estimation of the amplitude of the $Z^0$ contribution will be done with a cutoff regularisation, 
which is consistent if we restrict the study to the fermion self-energy. We will find that this contribution is proportional to 
the photon contribution, which will allow us to use the flows derived from DS
equations with dimensional regularisation. 

Before doing so, we need to discuss some crucial approximations regarding the resummation of the HECO effect of the $Z^0$-boson propagator. 
Due to the \h--\hb--$Z^0$ interactions \eqref{nonchiralHECO}, the $Z^0$ propagator will also need resummation over HECO loops, in a similar manner to the photon case. The inclusion of such a dressed/resummed propagator would algebraically complicate the above analysis, preventing an analytical treatment of the fixed point solution. Nonetheless, due to the smallness of the coupling of the interaction \eqref{nonchiralHECO}, which is proportional to $\sin\theta_W$, the experimental value of which is of order $\sin^2\theta_W \simeq 0.23$, a one-loop-DS-resummation correction to the dressed propagator would lead to a contribution to the relevant amplitude which would be one order of magnitude smaller than the corresponding one of the photon propagator. 

Moreover, when one considers $Z^0$ exchanges in DY processes, the bulk of the produced $Z^0$ bosons will be characterised by high momenta, which implies that such gauge bosons can be treated as massless in the relevant graphs. 
 In such a case, the dressed high-energy $Z^0$ propagator is expected to pick up correction terms similar in form to that of photon
\eqref{feynmanrules}, but with the $Z^0$-boson wave function renormalisation $\omega_{Z^0}$, and the other fixed-point parameters, assuming different values from the photon case. 

We shall consider the following effective Lagrangian with $R_\xi$ gauge-fixing for the $Z^0$ massive vector boson, with mass $M_Z$, which should be added to
\begin{align}\label{xiV}
\mathcal L_{Z^0} = -\frac{1}{2} \partial_\mu Z^0_\nu \Big(\partial^\mu Z^{0\nu} - \partial^\nu Z^{0\mu}\Big) + \frac{1}{2} M_Z^2 Z^0_\mu Z^{0\mu} - \frac{1}{2\xi_V} (\partial_\mu Z^{0\mu})^2\,,
\end{align}
where $\xi_V \equiv \frac{1}{\lambda_V}$ denotes the (vector meson) $R_\xi$ gauge fixing parameter, which does not need to be the same as that of the photon. 
From \eqref{xiV}, the $Z^0$ bare propagator reads (in momentum $p$ space):
\begin{align}
 \label{z0prop}\Delta^{Z^0}_{\mu\nu} = -\frac{i}{p^2- M_Z^2 + i\epsilon} 
 \Big(\eta_{\mu\nu} - (1 - \xi_V)\frac{p_\mu p_\nu}{p^2 - \xi_V\, M_Z^2}\Big)\,, \quad \epsilon \to 0^+\,.
\end{align}
In fact, in our treatment above, we found it convenient to use the Feynman gauge for the photon, having adopted the argument that this gauge is somehow a preferred one in other resummation schemes~\cite{pinched,pinched2}.
For the massive gauge boson at high energies, of interest to us in  this work,
given that the kinematic distribution in the production of the internal $Z^0$-bosons 
is such that the latter carry mostly high energies and momenta, 
it is convenient to use the $R_{\xi=0}$ 
(i.e.\ $\xi_V \to 0$) Landau gauge, which leads to a (approximately massless) $Z^0$-propagator before resummation, similar in form to 
that of photon:
\begin{align}\label{bareZprop}
 {\Delta^{Z^0}_{\mu\nu}}^{\rm high~energy} \simeq -\frac{i}{p^2 - M_Z^2 + i\epsilon} 
 \Big(\eta_{\mu\nu} - \frac{p_\mu p_\nu}{p^2}\Big)\,, \quad \epsilon \to 0^+\,.
\end{align}   
In the high-energy and high momentum limit $M_Z \ll E, |\vec p|$, the $M_Z$ mass is neglected in the denominator, 
which of course leads to  a photon-like propagator in the Landau gauge $\lambda_V \to +\infty$.
The one-loop-resummed DS dressing of this propagator by HECO loops will follow that of the photon in the corresponding gauge, leading to the form \eqref{ansatz} but with 
$\lambda_V \to +\infty$, which is the same as the bare propagator \eqref{bareZprop}.

In the Feynman gauge ($\xi_V=1/\lambda_V=1)$, in the high energy limit, the $Z^0$ boson behaves as a second photon, and one expects the dressing to lead to a factor in front of the transverse part of the propagator similar in form as in \eqref{feynmanrules}, but with the fixed value of the gauge-boson wave function renormalisation $\omega_{Z^0}$ being different from that of the photon, but of the same order of magnitude:
\begin{align}\label{bareZpropFeynGauge}
 {\Delta^{Z^0}_{\mu\nu}}^{\rm high~energy} \Big|^{\rm resum}_{\lambda_V=1} \simeq -\frac{i}{p^2 - M_Z^2 + i\epsilon} 
 \Big(\eta_{\mu\nu} + \omega_{Z^0} \,  \frac{p_\mu p_\nu}{p^2}\Big)\,, \quad \epsilon \to 0^+\,, 
\end{align}
where the mass $M_Z^2$ in the denominator  should be omitted in front of $p^2$ in the high energy limit, but is kept here for completeness and for future use. 
It is such fixed point expressions which, once included in the previous photon analysis, prevent an analytic solution of the corresponding fixed points, 
as already mentioned. 

Finally, in the unitary gauge $\xi_V \to +\infty$, for $M_Z \ne 0$, leads to the standard massive (Proca) $Z^0$ boson bare propagator 
\begin{align}\label{procaprop}
\Delta^{Z^0}_{\mu\nu} = -\frac{i}{p^2- M_Z^2 + i\epsilon} 
 \Big(\eta_{\mu\nu} - \frac{p_\mu p_\nu}{M_Z^2}\Big)\,, \quad \epsilon \to 0^+\,.
\end{align}
Unitary gauge means the absence of a gauge-fixing term, which is to be expected because for massive vector bosons, gauge fixing terms are not necessary.

We now remark that, as also discussed recently in \cite{Gallagher:2020ajd},
within the context of the SM, the unitary gauge is the default gauge for massive vector bosons, which acquire their masses via the Higgs mechanism.
In the unitary gauge, the propagator for the Goldstone scalar bosons arising from the 
spontaneously broken gauge symmetry procedure, which also have (gauge dependent) mass $M_Z$, vanishes, par contrast to the result in the Feynman gauge, where the massive vector boson and Goldstone boson propagators are similar. In fact, the gauge-dependent $Z^0$ propagator 
\eqref{z0prop} can be decomposed into 
gauge-parameter dependent and gauge-parameter independent parts as:
\begin{align}
 \Delta^{Z^0}_{\mu\nu} = -\frac{i}{p^2- M_Z^2 + i\epsilon} 
 \Big(\eta_{\mu\nu} - \frac{p_\mu p_\nu}{M_Z^2}\Big) - \frac{p_\mu p_\nu}{M_Z^2}\, \frac{i}{p^2 - \xi_Z m^2_Z + i\epsilon}\,, \quad \epsilon \to 0^+\,,
\end{align}
where the second ($\xi_V$ dependent) part on the right-hand side, which vanishes in the unitary gauge $\xi_V \to \infty$, is cancelled in an arbitrary $\xi_V$ gauge by the correspondiong neutral (in the case of the $Z^0$-boson) scalar Goldstone boson $h_Z$ propagator~\cite{Gallagher:2020ajd}. 

Performing the DS resummation in the context of the SM in the spontaneous-symmetry-breaking phase, that is combining photon and Proca massive vector boson propagators, is a complicated issue to be treated analytically, as already mentioned above. However, due to the relative smallness of the coupling of the HECO--$Z^0$ vertex compared to that of photon, we might expect that, on adopting the approximation of ignoring  resummation effects in the $Z^0$-boson propagator, thereby using the standard expression for the propagator \eqref{procaprop} of a massive free $Z^0$-Proca field, but retaining in full the above-described HECO-resummation effects in the photon--HECO vertex, photon-propagator, and $Z^0$-boson-resummed contributions to the HECO self-energy ({\it cf.}\ \eqref{s2s1}, \eqref{ghat}, below), the enhancement in the pertinent cross sections we find, which will be discussed in Section~\ref{sec:feyn}, will not be far (at least in order of magnitude) from the complete result. 

In the current work, we ignore $Z^0$--photon and $Z^0$--$Z^0$ fusion processes in the HECO production even at tree level, in similar spirit to what constitutes common practice in the current experimental searches. The reason is that at LHC energies the contribution of $Z^0/W^\pm$ bosons in protons is negligible with respect to other constituents, such as quarks, gluons and photons~\cite{Fornal:2018znf}. The situation changes when considering future hadron colliders at energies of ${\mathcal O}(100~\tev)$, where electroweak effects become more important and $Z^0 Z^0$ and $Z^0\gamma$ fusion may enhance the overall HECO production~\cite{Mangano:2016jyj}. This falls beyond the scope of this article.\footnote{We mention in passing that such contributions have been recently considered in the context of proton ultraperipheral  collisions (UPC), where it is pointed out that the inclusion of $Z^0$-exchange in PF processes of muon-pair production may, for large muon-pair invariant masses and large total transverse momentum, 
lead to enhancement of the pertinent cross section up to $20\%$~\cite{Godunov:2023myj}. We are not concerned here with UPC-specific processes.} 

%%%%%%%%%%%%%%%%%%%%%%%%%%%%%%%%%%%%%%%%%%%%%%%%
\subsection{HECO self-energy \label{sec:Hself}}

We now proceed to describe a few more analytical results that we can derive upon making the high-energy approximation for the exchanged $Z^0$-boson, which will be useful in producing the Feynman rules of the effective theory for HECO production to be used in Section~\ref{sec:feyn} for the UFO-model/\MAD implementation of the resummation effects. We concentrate on the HECO self-energy.
When the photon is the only gauge boson, the fermion self-energy for vanishing momentum is of order (taking into account \eqref{efflag}, \eqref{shiftcg})
\bea\label{photoncoupl}
\Sigma_1(0)=
\frac{ig^2}{1+\omega}\int\frac{d^4p}{(2\pi)^4}\gamma^\mu\frac{M+\slashed p}{p^2-M^2}\gamma^\nu\frac{\eta_{\mu\nu} 
+ \omega ~p_\mu p_\nu/p^2}{p^2}
=\frac{4 + \omega}{1+\omega} \frac{g^2M}{16\pi^2}\ln\left(\frac{\Lambda^2+M^2}{M^2}\right)~,
 \eea
where, in order to arrive at the final result on the right hand side, one passes first to a Euclidean formalism in momentum (Fourier) space, as appropriate for the implementation of an UV momentum cutoff $\Lambda$ in the momentum integrals, and the Feynman gauge is considered for the photon. 
Our aim here is to compare the contributions of the photon and the $Z^0$ boson to the HECO-fermion
self-energy; the relation between these contributions is independent on the regularisation method for the UV divergencies, as we shall verify explicitly below ({\it cf.} \eqref{s2s1}). From this point of view, the use of an UV momentum cut off $\Lambda$ is for the sake of simplicity and concreteness only.  
We now note that, in the expression for the photon propagator in \eqref{photoncoupl}, the corrections $\mathcal Z-1$  are neglected, since they are of order $\mathcal Z^\star \sim1$ and moreover they will not play an essential r\^ole in our arguments. We shall also argue below that the effects of the photon polarisation corrections can be cancelled by those of the $Z^0$ boson. 

Thus, only the fermion dressed mass $M$ should be  taken into account in the ``one-loop-like'' graph associated with the self-energy \eqref{photoncoupl}, which is sufficient for the present argument. Such an approximation allows for an analytic expression of the modified fixed point solution, when the $Z^0$-boson corrections are taken into account. 

Indeed, if the HECO fermion also couples to a $Z^0$ boson with mass $M_Z$,
the DS equation for the fermion self-energy 
contains an additional contribution $\Sigma_2$, where the dressed fermion propagator is used inside the loop. 
We consider the expression 
\eqref{bareZpropFeynGauge}
for the 
one-loop resummed $Z^0$-boson propagator, in the high energy limit and in Feynman gauge, which leads to (where, again, it is understood that in order to arrive at the result on the right-hand side, the integration momenta and the gauge boson exchange  propagators should be analytically continued to a Euclidean Fourier space, in the presence of $\Lambda$):
\begin{align}\label{vector}
\Sigma_2(0)&=
\frac{ig'^2}{1+\omega_{Z^0}}\int\frac{d^4p}{(2\pi)^4}\gamma^\mu\frac{M+\slashed p}{p^2-M^2}\gamma^\nu\frac{\eta_{\mu\nu} +\omega_{Z^0} p_\mu p_\nu/p^2}{p^2-M_Z^2} \nonumber \\
&=\frac{4+ \omega_{Z^0}}{(1+\omega_{Z^0})16\pi^2}\frac{g'^2M}{M^2-M_Z^2}\times\left(M^2\ln\left(\frac{\Lambda^2+M^2}{M^2}\right)-M_Z^2\ln\left(\frac{\Lambda^2+M_Z^2}{M_Z^2}\right)\right)~, 
\end{align}
where $g'$ is the appropriate coupling determined from Eq.~\eqref{nonchiralHECO}.
As already mentioned, we are interested in the high-energy limit, that is a situation where $M_Z\ll M<\Lambda$, in which case the $Z^0$ boson self-energy is approximately proportional to the photon self-energy \eqref{photoncoupl}: 
\be\label{s2s1}
\Sigma_2(0)\simeq \frac{(4+\omega_{Z^0}) \, (1 + \omega)}{(1+\omega_{Z^0})\, (4+\omega)} 
\frac{g'^2}{g^2}\Sigma_1(0) \equiv \mathcal{A} \, \frac{g'^2}{g^2}\Sigma_1(0)~.
\ee
This implies that the previous calculation of the fermion self-energy, assuming only photon exchange, can be extended to include the $Z^0$ boson contributions by employing the effective replacement:
\begin{align}\label{ghat}
g^2 \to \hat g^2 &\equiv g^2+ {\cal A} \,g'^2~.
\end{align}
Given the different processes we consider and the similarities between the propagators of the high-energy $Z^0$-boson and the photon,
in the numerical approach we will consider the approximation ${\cal A}\sim1$. 
Specifically, in our subsequent analysis in Section~\ref{sec:feyn}, we shall use for concreteness the value:
\begin{align}\label{ghatA3}
\mathcal A = 3/4~.
\end{align}
This follows from the approximation of using bare exchanged-photon propagator in the Feynman gauge, {\it i.e.} $\omega=0$, and bare exchanged-$Z^0$ boson propagator in the $R_{\xi=0}$ gauge \eqref{bareZprop} in the 
self-energy loops (which would result in an expression like \eqref{vector}, but with $
\omega_{Z^0} $ in the numerator of the integrand being replaced by $-1$ and  $\omega_{Z^0}=0$ in the denominator).

The photon self-energy is not modified though, under the above approximations. Hence, the analogue of the flows (\ref{ZomegaM}) involve now the two coupling constants
$g^2$ and $\hat g^2$
\bea
\mathcal Z&=&1+\frac{\hat g^2}{8\pi^2}\ln\left(\frac{\mathcal Z\,k}{M}\right)~, \nonumber \\
\mathcal Z\omega&=&\frac{g^2}{6\pi^2}\ln\left(\frac{\mathcal Z\,k}{M}\right)~,\nn
\mathcal Z\left(1-\frac{m}{M}\right)&=&\frac{\hat g^2}{8\pi^2}\frac{4+\omega}{1+\omega}\ln\left(\frac{\mathcal Z\,k}{M}\right)~.
\eea
Hence we find here instead of Eq.~(\ref{k})
\be
\frac{k}{m}=\frac{(3+4\eta)\mathcal Z-4\eta}{9(\hat{ \mathcal Z}_+-\mathcal Z)(\mathcal Z-\hat{\mathcal Z}_-)}~\exp\left(\frac{8\pi^2}{\hat g^2}(\mathcal Z-1)\right)~,
\ee
where $\eta\equiv g^2/\hat g^2<1$ and
\be
\hat{\mathcal Z}_\pm=\frac{2}{9}(3+\eta)\left(1\pm\sqrt{1-\frac{9\eta}{(3+\eta)^2}}\right)~.
\ee
Note that, for any value of $\eta$, both $\hat{\mathcal Z}_+$ and $\hat{\mathcal Z}_-$ are real and $\hat{\mathcal Z}_+>1$. 
For the actual range $0<\eta<1$ though, we have $\hat{\mathcal Z}_-<1$, such that the relevant regime with $k>0$ is $1\le \mathcal Z\le \hat{\mathcal Z}_+$.
The situation is thus similar to the case with the photon only, where the UV fixed point is instead 
\bea\label{mheco}
\hat{\mathcal Z}^\star&=&\hat{\mathcal Z}_+~, \nonumber \\
\hat\omega^\star&=&\frac{4}{3}\eta\left(1-\frac{1}{\hat{ \mathcal Z}^\star}\right)~, \nn
\left(\frac{M}{k}\right)^\star&=&\hat{\mathcal Z}^\star\exp\left(-\frac{2\pi}{\hat\alpha^\star}(\hat{\mathcal Z}^\star-1)\right)~,
\eea
where $\hat\alpha^\star = \frac{g^2}{4\pi\,\eta} = \frac{\hat g^2}{4\pi}$. As in the case of pure QED, discussed previously in Sections~\ref{sec:rc} and~\ref{sec:rm}, 
the effective theory describing the tree-level DY or PF processes, contains the rescaled coupling by the factor $1/\sqrt{1 + \omega}$. So the value of the effective fine-structure parameter at the fixed point in this case, taking into account the $Z^0$-effects, is:
\be\label{alphastar2}
\hat\alpha_\text{eff}^\star=\frac{\hat g^2/4\pi}{1+\hat\omega^\star}~.
\ee
It can be seen numerically that, for a fixed scale $k$, the physical mass $M(k)/Z(k)$ increases when $g'$ increases 
(or equivalently $\eta$ decreases), which is expected from the fact that a stronger interaction induces a larger correction to the mass.

 Notice that, if we identified the UV fixed-point scale with that of the UV cutoff $\Lambda$ (which, we stress again, in our approach does not go to $\infty$, but it denotes the scale of new physics at the collider searches at hand), then the aforementioned fixed point value of $M/\mathcal Z$ in the presence of $Z^0-\overline{\mathcal H}-\mathcal H$ interactions:
\begin{align}\label{mhat}
\hat{\mathcal M}^\star \equiv \mathcal M (k=\Lambda)\,, 
\end{align}
differs from that of the previous Section~\ref{sec:rm}, \eqref{mhgonly}, 
as follows from the fact that the values of the pertinent UV fixed point parameters \eqref{mheco} and \eqref{alphastar2}  differ from those (\eqref{omstasrdef}, \eqref{fixedpoint}) of the corresponding UV fixed point in the (only-$\gamma\rm )-\overline{\mathcal H}-\mathcal H$ coupling. Such a different value is denoted by a hat over the fixed-point mass symbol. This is an important point with physical significance that we should bear in mind, when attempting to extract mass bounds from HECO searches at colliders upon the application of our resummation scheme.

Before closing the section, we would like to remark, for completeness, that 
in the case of more general chiral couplings $c_L \ne c _R$ in Eq.~\eqref{chiralHECO}, the situation becomes more complicated, as one may have both constructive and destructive interferences to the self-energy from the contribution of the $Z^0$-boson (or other new neutral bosons, much lighter than the HECO). For instance, in the case of purely axial couplings, $c_L=-c_R$, instead of Eq.~\eqref{nonchiralHECO}, one obtains a  purely axial interaction: 
\begin{align}\label{axialHECO}
\mathcal L^A_\text{int} = \frac{e}{2\, \sin(2\theta_W)}\,c_L \hb \, \gamma^\mu\, Z^0_\mu  \gamma^5 \, \h \,.
\end{align} 
As a result of the $\gamma^\mu\gamma^5$ Lorentz structure in the appropriate vertex, the one-loop DS equation for the $Z^0$-boson self-energy contribution\footnote{For brevity and concreteness in our discussion, we still assume in this case only a Dirac mass $M$ for the HECO fermion. Extension to incorporate also a chiral mass $M^\prime \gamma^5$ for the particle is possible in our resummation scheme, but it will not be studied here.} will now have an overall minus sign on its right-hand side, i.e.\ the result will be obtained by the replacement $M \to -M$ in Eq.~\eqref{vector}, which will also imply an overall minus sign in the analogue of Eq.~\eqref{s2s1}. 
Thus, the only difference between the one-loop self-energies in the axial and vector cases is an overall sign in front of the quantity $M$ in the relevant    expressions.
This, in turn, will lead to a combined HECO self-energy corresponding to an effective coupling in the purely axial ($A$) case ({\it cf}.\  Eq.~\eqref{ghat}):
\begin{align}\label{ghat2}
g^2 \to \hat g^2 &\equiv g^2 - {\cal A}g_A'^2~,
\end{align}
where $g_A'$ is the coupling corresponding to the axial interaction \eqref{axialHECO}.  This signifies a destructive interference in the self-energy of the $Z^0$ boson contribution in the resummation scheme at hand, in contrast to the constructive interference of the vector coupling \eqref{ghat}. When  both axial ($g_A'$) and vector ($g_V'$) couplings are present, the right-hand-side of Eq.~\eqref{ghat2} is modified by the addition of the term $+{\cal A}g_V'^2$.

%%%%%%%%%%%%%%%%%%%%%%%%%%%%%%%%%%%%%%%%%%%%%%%%
%%%%%%%%%%%%%%%%%%%%%%%%%%%%%%%%%%%%%%%%%%%%%%%%
\section{\MAD implementation and cross-section estimation}\label{sec:feyn}

The implementation of the above-described resummation effects has been carried out specifically for Universal Feynrules~\cite{Alloul:2013bka} Output (UFO) models~\cite{ufo}. These models are designed to be compatible with Monte Carlo event generators such as \madamc~3.5.1~\cite{madgraph}, which is the tool employed in this study. 

The primary goal is to simulate the production mechanisms of Drell-Yan and photon fusion  involving spin-\hf HECOs at the LHC using the resummation scheme described in the previous section. For the case of DY production, both $\gamma$-only and $\gamma/Z^0$ exchanges have been considered in the implementation.
Two distinct models have been developed: 
\begin{description}
\item[$\gamma$ exchange UFO model] It takes into account the sole contribution from $\gamma$ exchange, which is suitable for simulating both DY and PF processes. It uses the Feynman rules \eqref{feynmanrules}, stemming from the effective Lagrangian \eqref{efflag}, at the UV fixed point \eqref{omstasrdef}, \eqref{effcoupl} and \eqref{MLambda}. 
\item[$\gamma/Z^0$ exchange UFO model] This model includes the additional exchange of the $Z^0$ boson in DY production. It makes use of the pertinent Feynman rules at the UV fixed point \eqref{ghat}, \eqref{ghatA3}, \eqref{mheco}, \eqref{alphastar2}, in the non-chiral vector case \eqref{nonchiralHECO}, in which we restrict our attention here.
\end{description}
By incorporating the resummation effects into the UFO models, we are able to provide reliable predictions for the production mechanisms under study. 
The simulations using the implemented models allow for the study of various observables related to DY and PF processes involving spin-\hf HECOs at the LHC and the effects of the resummation on the predicted outcomes. These UFO models are meant to become a tool for comparison of experimental  results with theoretical benchmarks. A summary of the Feynmam rules used and the validation procedure and results are given in Appendices~\ref{feynman_appendix} and~\ref{validation_appendix}, respectively. The UFO models are validated by comparing the simulation-estimated cross section values against the outcome of analytical calculations using \Math.

In addition to the standard input parameters such as centre-of-mass energy and parton distribution functions (PDFs), the UFO models introduced in this study incorporates two new parameters:
\begin{itemize}%[nosep]
\item The multiplicity of the charge, denoted by $n$, which appears in the coupling definition $g = n e$, with $e$ being the electron charge. The HECO charge is also denoted by $Q=ne$.
\item The cutoff energy scale parameter, denoted by $\Lambda$, defined in Eq.~\eqref{Lambda}. 
\end{itemize} 

Both the multiplicity and the cutoff parameter have a direct impact on the HECO mass and, consequently, affect the calculated cross-section value.
Furthermore, for the specific centre-of-mass energies considered in this study, i.e., those at the LHC, the fine-structure constant is set $\alpha_\text{EM}=1/127.94$, i.e.\ to the value pertinent for the $Z^0$-boson mass scale. We stress here that the choice of this value affects significantly the absolute cross-section value of the results. 

Tables~\ref{tabDYtreelevsresum},~\ref{tabDYZ0treelevsresum} and~\ref{tabPFtreelevsresum} provide a comparison between the cross-section values obtained at the tree level and after incorporating the resummation effects for both the Drell-Yan and photon-fusion processes. The calculations are performed considering $pp$ collisions at $\sqrt{s}=13~\tev$.
For the DY process, the \texttt{NNPDF23}~\cite{NPDF} PDF is utilised, while for the PF process, the \texttt{LUXqed17}~\cite{luxqed} PDF is employed. These PDF choices are used in conjunction with the UFO model to calculate the cross-section values.
Upon comparing the cross-section values before and after including resummation effects, we see that for the same mass --- bare for tree level and ${\cal M}(\Lambda)$ for the resummation --- the cross section is of the same order of magnitude, thus validating the UFO models and the underlying calculations. Concerning the effect of DS resummation, it is observed that there is a significant increase in the cross-section value. 
In addition, the cross-section values related to the PF process turn out to be higher than those of the DY process at the LHC energies and for this range of HECO masses. This behavior arises from the cross-section dependence which is $\propto n^2$ for DY, whereas it is $\propto n^4$ for PF.

\begin{table}[ht]
\caption{Cross-section comparison for HECO production in $pp$ collisions at $\sqrt{s}=13~\tev$ via the Drell-Yan ($\gamma-$only) process between tree level and after resummation with $\Lambda= 2~\tev$. The HECO mass is also listed.}
\label{tabDYtreelevsresum}
\centering
\begin{tabular}{ >{\centering}p{1.1cm}>{\centering}p{3.3cm}>{\centering}p{3.3cm}>{\centering\arraybackslash}p{2.3cm}  }
\hline\hline
\multicolumn{4}{ c }{DY $p p \to \gamma \to \h\hb$ @ $\sqrt{s}=13~\tev$, $\Lambda= 2~\tev$} \\
\hline
$Q~(e)$ & $\sigma_\text{tree-level}$ (fb) & $\sigma_\text{resum}$ (fb) & $M$ (\tev) \\
\hline
20 & 775 & 1692 & 0.507 \\
60 & 4.959 & 10.79 & 1.717 \\
100 & 5.949 & 12.95  & 1.893 \\
140 & 9.134 & 19.86  & 1.945 \\
180 & 13.67  & 29.71  & 1.966 \\
220 & 19.31  & 42.08  & 1.977 \\
\hline\hline
\end{tabular} 
\end{table}

\begin{table}[ht]
\caption{Cross-section comparison for HECO production in $pp$ collisions at $\sqrt{s}=13~\tev$ via the Drell-Yan (including $Z^0$) process at tree level and after resummation with $\Lambda= 2~\tev$. The HECO mass is also listed.}
\label{tabDYZ0treelevsresum}
\centering
\begin{tabular}{ >{\centering}p{1.1cm}>{\centering}p{3.3cm}>{\centering}p{3.3cm}>{\centering\arraybackslash}p{2.3cm}  }
\hline\hline
\multicolumn{4}{ c }{DY $p p \to \gamma/Z^0 \to \h\hb$  @ $\sqrt{s}=13~\tev$, $\Lambda= 2~\tev$}  \\
\hline
$Q~(e)$ & $\sigma_\text{tree-level}$ (fb) & $\sigma_\text{resum}$ (fb) & $M$ (\tev) \\
\hline
20 & 101.4 & 211.8 & 0.758 \\
60 & 3.125 & 6.527 & 1.796 \\
100 & 4.722 & 9.835 & 1.924 \\
140 & 7.752 & 16.25  & 1.961 \\
180 & 11.95  & 24.92  & 1.976 \\
220 & 17.22  & 35.94  & 1.984 \\
\hline\hline
\end{tabular} 
\end{table}

\begin{table}[ht]
\caption{Cross-section comparison for HECO production in $pp$ collisions at $\sqrt{s}=13~\tev$ via the photon-fusion process at tree level and after resummation with $\Lambda= 2~\tev$. The HECO mass is also listed. The PDF \texttt{LUXqed17} is used.}
\label{tabPFtreelevsresum}
\centering
\begin{tabular}{ >{\centering}p{1.1cm}>{\centering}p{3.3cm}>{\centering}p{3.3cm}>{\centering\arraybackslash}p{2.3cm}  }
\hline\hline
\multicolumn{4}{ c }{PF $p p \to \h\hb$  @ $\sqrt{s}=13$ TeV, $\Lambda= 2~\tev$}  \\
\hline
$Q~(e)$ & $\sigma_\text{tree-level}$ (fb) & $\sigma_\text{resum}$ (fb) & $M$ (\tev) \\
\hline
20 & 1.321 $\times 10^{4}$ & 6.271 $\times 10^{4}$ & 0.507 \\
60 & 8.466 $\times 10^{2}$ & 4.025 $\times 10^{3}$ & 1.717 \\
100 & 2.895 $\times 10^{3}$ & 1.372 $\times 10^{4}$ & 1.893 \\
140 & 8.753 $\times 10^{3}$ & 4.170 $\times 10^{4}$ & 1.945 \\
180 & 2.175 $\times 10^{4}$ & 1.030 $\times 10^{5}$ & 1.966 \\
220 & 4.612 $\times 10^{4}$ & 2.184 $\times 10^{5}$ & 1.977 \\
\hline
\end{tabular} 
\end{table}

Another interesting point, seen when comparing the Tables~\ref{tabDYtreelevsresum} and~\ref{tabPFtreelevsresum} with Table~\ref{tabDYZ0treelevsresum}, is that the HECO mass for a specific value of charge and cutoff scale depends on the inclusion of the $Z^0$ in the vertex. As discussed in Section~\ref{sec:zb} ({\it cf.} 
Eqs.~\eqref{mheco}, \eqref{mhat}), 
the inclusion of $Z^0$ in the vertex leads to a higher HECO mass $M$ compared to the case where only \h--\hb--$\gamma$ interactions are considered. We reiterate here that the UFO model with \h--\hb--$\gamma$ is used for resummation in DY process with $\gamma$-only (Table~\ref{tabDYtreelevsresum}) and in PF (Table~\ref{tabPFtreelevsresum}, whereas the \h--\hb--$\gamma/Z^0$ UFO model is deployed for the DY with $\gamma/Z^0$ exchange (Table~\ref{tabDYZ0treelevsresum}). 

\begin{figure}   
    \centering
    \includegraphics[width=0.5\textwidth] {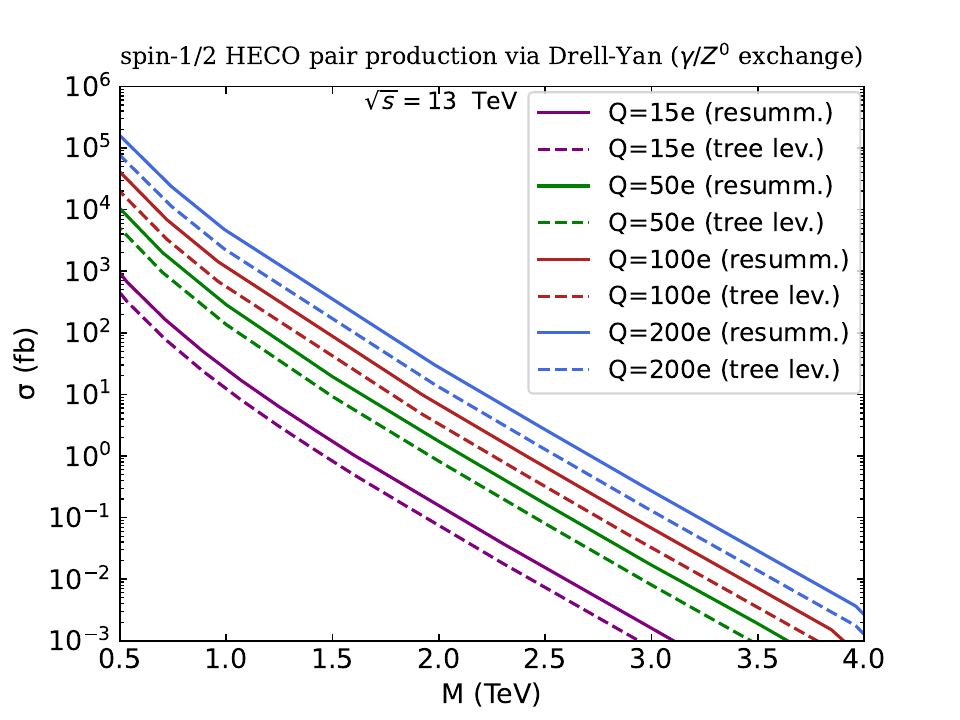}\hfill
    \includegraphics[width=0.5\textwidth] {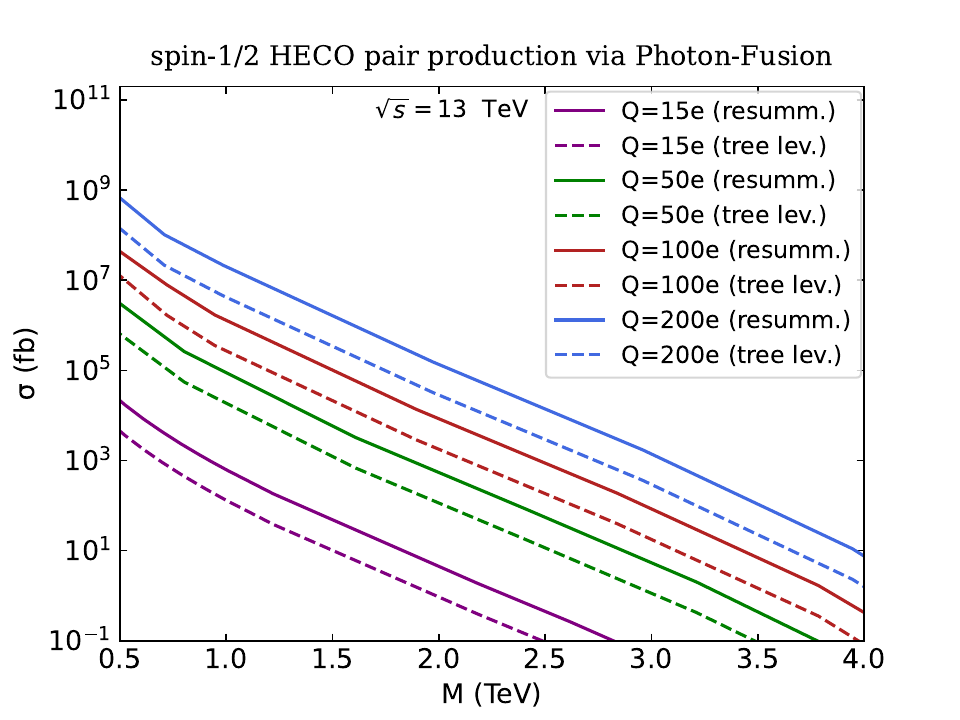}
    \caption{Comparison of cross section values obtained after (solid line) and before (dashed-line) resummation as a function of the HECO mass $M$ for electric charges $Q=15e$, $50e$, $100e$ and $200e$, demonstrating the significant effect of the resummation by increasing the corresponding cross sections.} 
    \label{fig:aftervsbefore}
\end{figure}

The aforementioned points are also demonstrated in Figure~\ref{fig:aftervsbefore} where the cross section versus the HECO mass $M$ is drawn before and after resummation at $\sqrt{s}=13~\tev$ for various charge values. We observe that the production becomes more abundant with resummation by a factor of $\sim 2.1$ and $\sim 4.75$ for DY and PF, respectively. The stronger impact of the resummation on the PF cross section when compared with the DY process is due to the presence of two vertices involving a HECO in the former as opposed to one vertex in the latter, as shown in the related diagrams in Figure~\ref{fig:process}.

\begin{figure}[ht]
\includegraphics[width=.5\textwidth ]{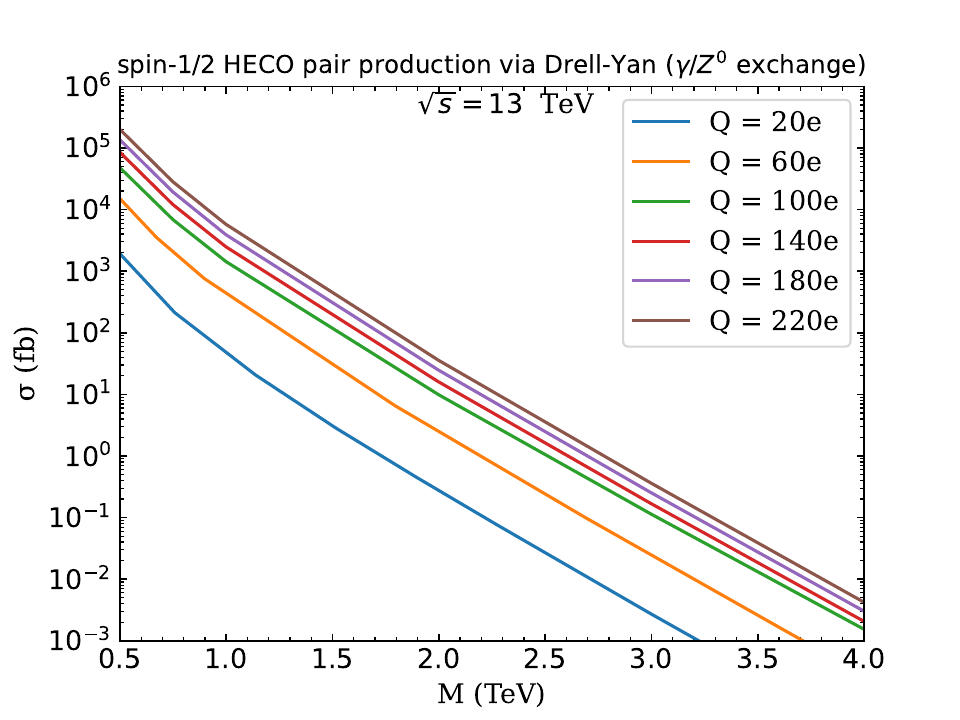}\hfill
\includegraphics[width=.5\textwidth]{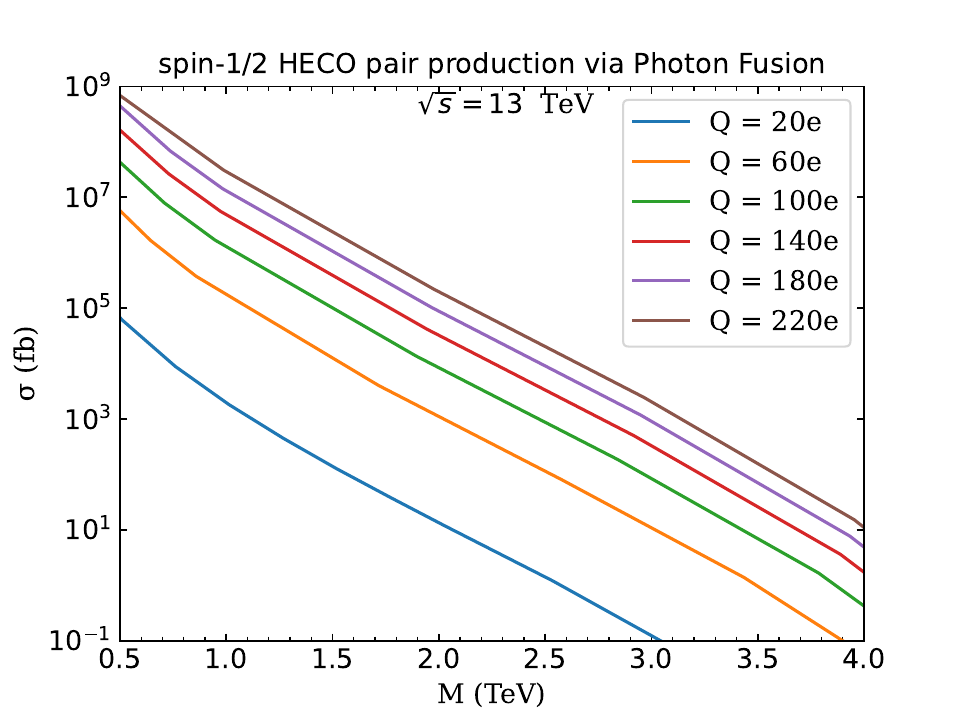}
\\[\smallskipamount]
\includegraphics[width=.5\textwidth]{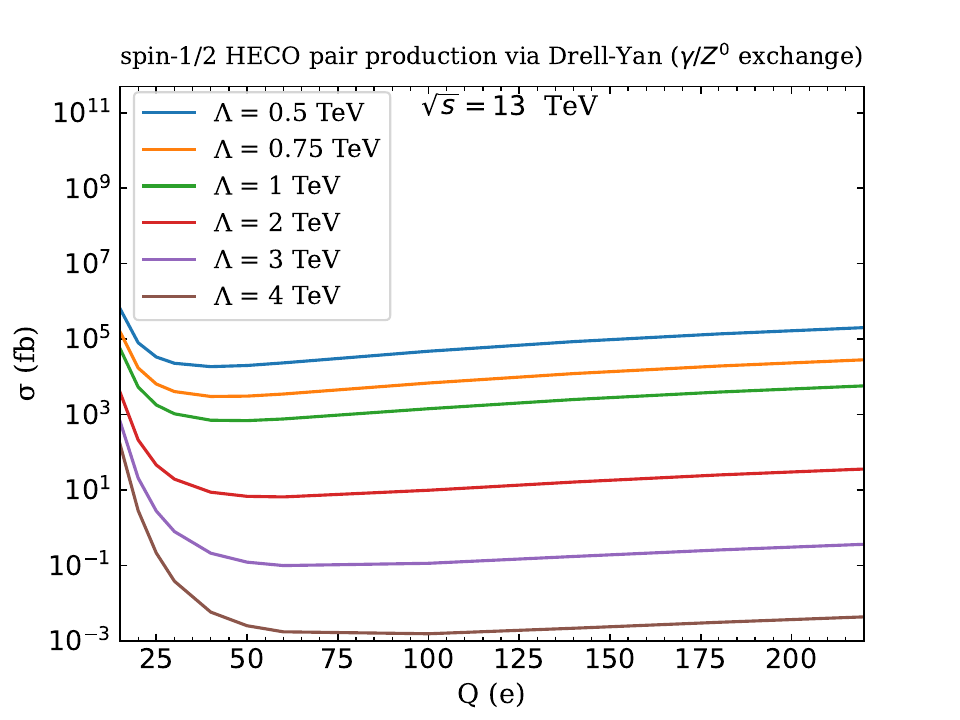}\hfill 
\includegraphics[width=.5\textwidth]{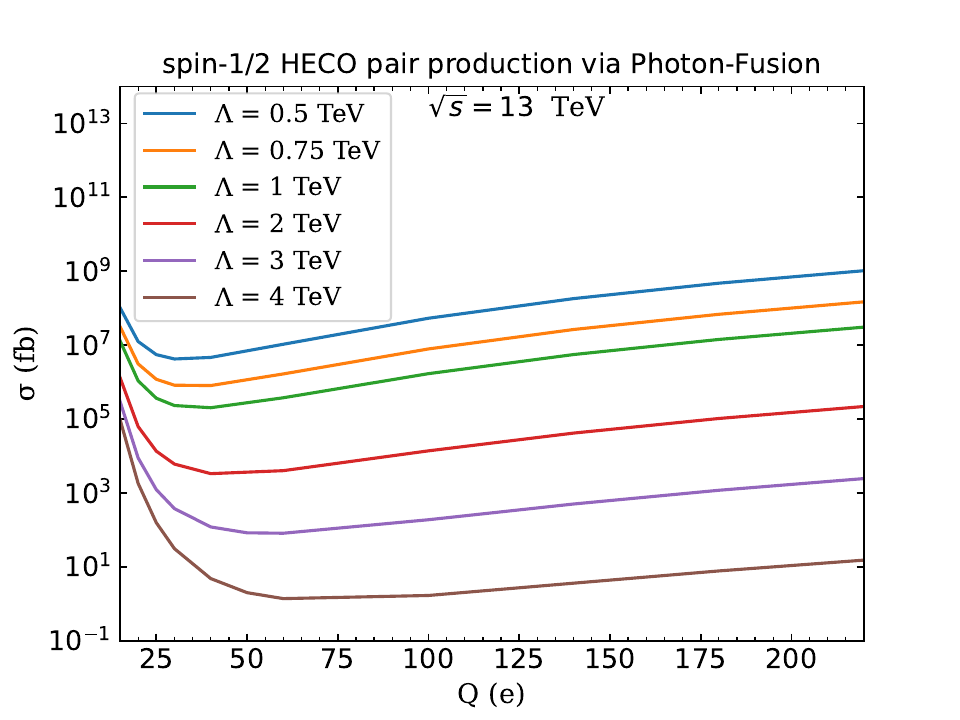}
\includegraphics[width=.5\textwidth]{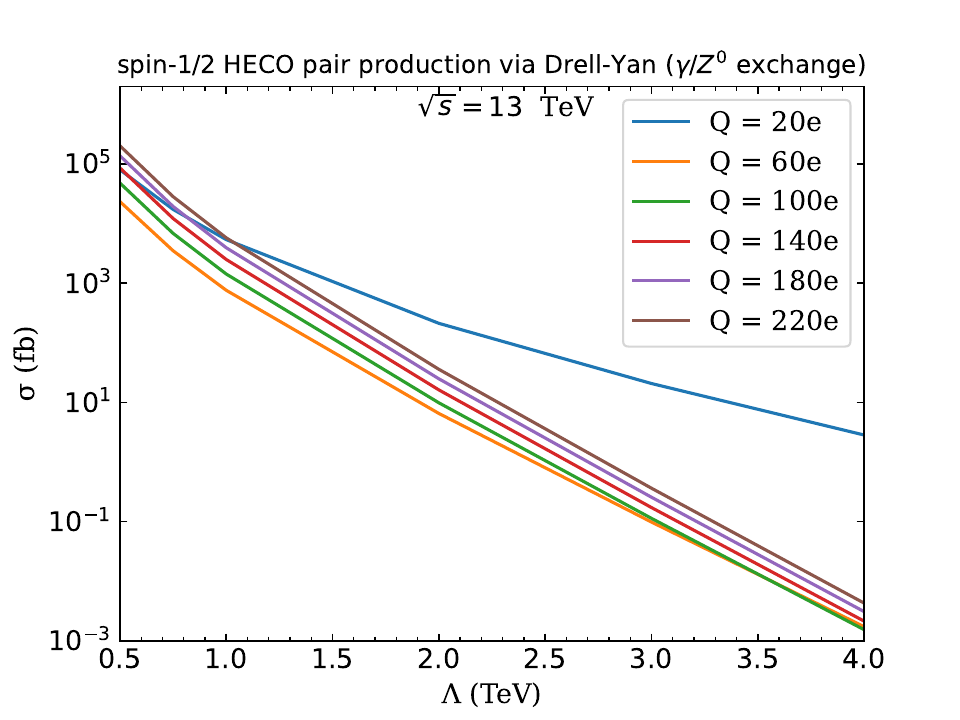}\hfill  
\includegraphics[width=.5\textwidth]{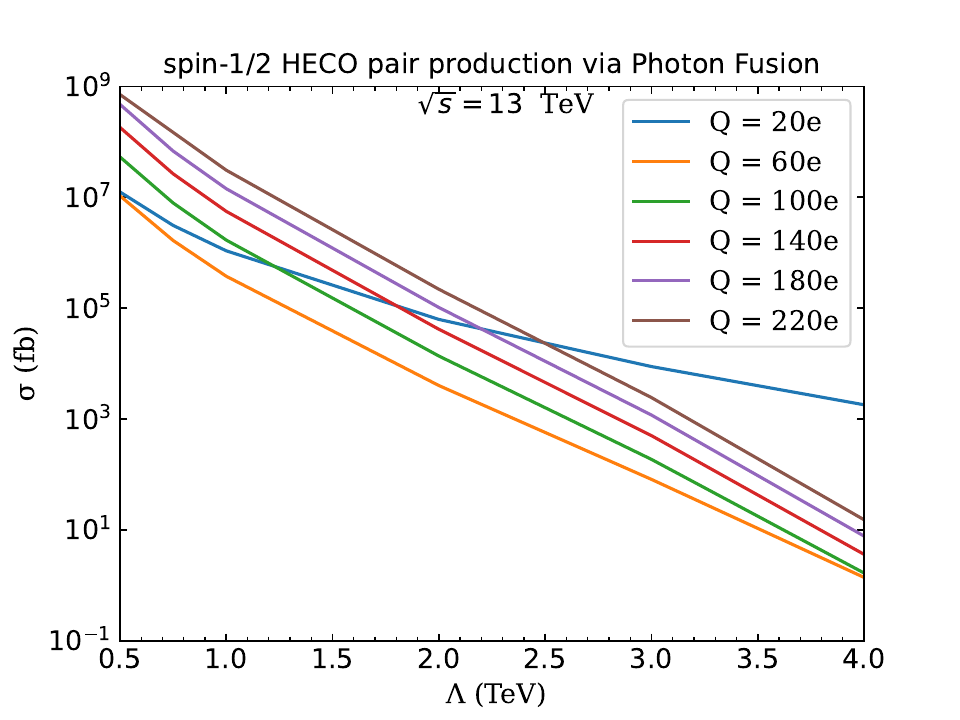}
\\[\smallskipamount]
    \caption{Cross-section values for both DY with $\gamma/Z^0$ exchange (left) and PF (right) after resummation. The HECO pair cross section production at $\sqrt{s}=13~\tev$ is drawn as a function of: (top) the HECO mass $M$ for various charge values; (center) the charge $Q$ for different cutoff $\Lambda$ values; and (bottom) the cutoff $\Lambda$ for various $Q$ values.}
    \label{fig:DY_plots}
\end{figure}

Figure~\ref{fig:DY_plots} shows the cross-section values for DY (left column) and PF (right column) processes versus different parameters. As expected, the cross section consistently increases as $Q$ becomes larger for the same HECO mass $M$. The behaviour changes when varying the charge: the cross section rapidly drops with increasing $Q$ till a $\Lambda$-dependent charge value is reached, beyond which cross section slightly increases. This non-monotonous behaviour reflects the non-linear relation of $\Lambda$ with charge given in analytical form in Eq.~\eqref{MLambda}. The increase in $n$, i.e.\ $Q$, leads to a significant contribution from the HECO mass term, causing a decrease in the production rate. This also explains the larger cross section of $Q=20e$ compared to higher charges for a cutoff scale larger than a critical value of $\Lambda$. We note again here the consistently more abundant production characterising PF when compared to the DY process at the considered collision energy of 13~\tev.

While here we mainly focus on $pp$ collisions, motivated by the recent searches conducted by LHC experiments such as ATLAS~\cite{ATLAS:2023zxo} and MoEDAL~\cite{MoEDAL:2021mpi}, it is possible to extend the analysis to other colliding particles. By changing the type of collisions to $e^+e^-$ or $\mu^+\mu^-$, for instance, the potential for discovering HECOs at future lepton colliders can be explored using this model. This flexibility allows for a broader investigation of HECOs and their detection prospects in various collider environments.

%%%%%%%%%%%%%%%%%%%%%%%%%%%%%%%%%%%%%%%%%%%%%%%%
%%%%%%%%%%%%%%%%%%%%%%%%%%%%%%%%%%%%%%%%%%%%%%%%
\section{Improved Mass Limits} \label{sec:limits} 

In this section, we re-interpret HECO search results performed by the ATLAS~\cite{ATLAS:2008xda} and MoEDAL~\cite{MoEDAL:2009jwa} experiments at the LHC in view of the (more accurate) predictions for the production cross sections obtained in Section~\ref{sec:feyn} that include resummation effects.  

ATLAS~\cite{ATLAS:2008xda}, one of the largest main experiments at the LHC, is a multi-purpose detection system optimised to directly detect known SM particles. Despite that, it has also proved to be sensitive to HIPs, such as HECOs, in analyses of data recorded at 8~\tev~\cite{ATLAS:2015tyu} and 13~\tev~\cite{ATLAS:2019wkg,ATLAS:2023esy} $pp$ collisions. Such searches rely mostly on high-ionisation signals from the transition radiation tracker~\cite{Mitsou:2003rp} and the electromagnetic calorimeter~\cite{ATLAS:2010blk}. The very recent analysis on 138~\ifb of data collected between 2015 and 2018 considered for the first time the PF HECO production process and excluded HECO charges $20 \leq |Q| \leq 100$, for masses up to 3.1~\tev~\cite{ATLAS:2023esy}. The mass limits of these three ATLAS analysis are listed in Table~\ref{tab:limits} in the columns titled ``LO'' (leading order).

\begin{table*}[ht]
\caption{95\% CL experimental mass limits for spin-\hf HECOs with (DS) and without (LO) Dyson-Schwinger resummation techniques to calculate production cross sections. Reported bounds are not necessarily extracted from the same dataset.}
\label{tab:limits}
\centering
\begin{tabular}{ c c c c c c c c c c c }
\hline\hline
\multicolumn{11}{ c }{Experimental lower limits at 95\% CL on spin-\hf HECO mass (\tev)}\\
\hline
\multirow{2}{2.2cm}{Experiment/ energy} & \multirow{2}{*}{$Q~(e)$} & \phantom{99} & \multicolumn{2}{c}{DY $\gamma$ exchange} & \phantom{99} & \multicolumn{2}{c}{DY $\gamma$/$Z^0$ exchange} & \phantom{99} &\multicolumn{2}{c}{$\gamma\gamma$ fusion} \\ 
\cline{4-5}\cline{7-8}\cline{10-11}
 & & & LO & DS & & LO & DS & & LO & DS\\
\hline
\multirow{11}{2.2cm}{MoEDAL~\cite{MoEDAL:2021mpi} $\sqrt{s}=8~\tev$} & 15 && 0.18 & 0.24 && 0.17 & 0.24 && -- & -- \\
& 20   && 0.28 & 0.36 && 0.31 & 0.36 && -- & -- \\
& 25   && 0.44 & 0.55 && 0.44 & 0.53 && -- & -- \\
& 50   && 0.78 & 0.88 && 0.78 & 0.87 && -- & -- \\
& 75   && 0.78 & 0.88 && 0.78 & 0.84 && -- & -- \\
& 100  && 0.73 & 0.84 && 0.71 & 0.80 && -- & -- \\
& 125  && 0.66 & 0.75 && 0.64 & 0.72 && -- & -- \\
& 130  && 0.64 & 0.74 && 0.62 & 0.70 && -- & -- \\
& 140  && 0.58 & 0.68 && 0.62 & 0.69 && -- & -- \\
& 145  && 0.52 & 0.66 && 0.51 & 0.60 && -- & -- \\
& 150  && 0.50 & 0.63 && 0.58 & 0.66 && -- & -- \\
\hline
\multirow{5}{2.2cm}{ATLAS~\cite{ATLAS:2015tyu} $\sqrt{s}=8~\tev$} & 10 && 0.78 & 0.78\footnotemark[1] && -- & -- && -- & --  \\
& 20 && 1.05 & 1.14 && -- & -- && -- & -- \\
& 40 && 1.16 & 1.25 && -- & -- && -- & -- \\
& 60 && 1.07 & 1.15 && -- & -- && -- & -- \\
\hline
\multirow{5}{2.2cm}{ATLAS~\cite{ATLAS:2019wkg,ATLAS:2023esy} $\sqrt{s}=13~\tev$} & 20 && 1.83 & 2.02 && 1.8 & 1.9 &  & 2.5 &  2.7  \\
& 40   && 2.05 & 2.22 && 2.2 & 2.3 && 3.1 & 3.4 \\
& 60   && 2.00 & 2.18 && 2.2 & 2.4 && 3.1 & 3.4 \\
& 80   && 1.86 & 2.02 && 2.1 & 2.2 && 3.0 & 3.0\footnotemark[2] \\
& 100  && 1.65 & 1.80 && 1.9 & 2.1 && 2.5 & 2.5\footnotemark[2] \\
\hline\hline
\end{tabular}
\footnotetext[1]{Resummation is valid for $Q\gtrsim11e$, so there is no change in the mass limit for $Q=10e$.}
\footnotetext[2]{There is no experimental sensitivity for HECO masses higher than this value.}
\end{table*}

MoEDAL~\cite{MoEDAL:2009jwa,dcmp}, on the other hand, is an experiment specialised on HIPs, including magnetic monopoles and HECOs; the deployment of Nuclear Track Detectors provides sensitivity to the latter. The first analysis for HECOs was carried out using a prototype array of Makrofol plastic sheets exposed to 8~\tev $pp$ collisions, considering DY production with and without $Z^0$-boson exchange. The resulting upper limits placed on HECO production cross section vary from $\sim30$~fb to 70~pb, for electric charges in the range $15e$ to $175e$ and masses from 110~\gev to 1020~\gev~\cite{MoEDAL:2021mpi}; the latter listed in the respective columns ``LO'' of Table~\ref{tab:limits}.

All aforementioned searches derived lower limits on HECO masses by confronting the experimentally set upper limits of HECO pair production cross sections with the theoretical predictions obtained by calculations at \emph{tree level}. These cross-section upper limits only depend on the assumed angular distribution of the simulated particle production.\footnote{An illustration of such angular distribution dependence on the production mechanism ---Drell-Yan and $\gamma\gamma$ fusion--- is given in Ref.~\cite{baines} for magnetic monopoles.} Given that the resummation only rescales the total cross section without affecting the differential cross section with respect to kinematic variables, such as the rapidity and the transverse momentum, the cross-section upper limits can be safely compared with the cross section after resummation for each  process, aiming at extracting mass limits. 

Improved and \emph{reliable} mass limits are provided by our DS resummation treatment discussed in this work. These limits are presented in the columns ``DS'' of Table~\ref{tab:limits} for various ATLAS and MoEDAL analyses and production processes using the cross sections calculated in Section~\ref{sec:feyn}. We observe that the updated limits are consistently more stringent than the originally obtained by the experimental collaborations, in agreement with the higher cross sections obtained when resummation effects are added compared to the tree-level production rates, as demonstrated in Section~\ref{sec:feyn}. The increase in the mass limits spans a wide range of values up to 30\%, as the cross-section upper limits show an irregular pattern depending on the HECO mass and the experiment. It is reminded that results for the same experiment and $\sqrt{s}$ may not necessarily correspond to the same dataset size, hence they cannot be compared with each other. 

A summary of the spin-\hf HECO mass limits obtained in this work versus the HECO charge is given in the graph of Figure~\ref{fig:hecos-resum}. These results include the DS resummation effects and are based on the re-interpretation of HECO production-rate constraints obtained by the ATLAS and MoEDAL. ATLAS has better limits for charges up to $100e$, while MoEDAL takes over at higher charges. A new analysis of the full Run~2 MoEDAL detector, comprising the full NTD detector, is expected to be released soon constraining even higher electric charges~\cite{MoEDAL:2023ost}. Given that the mass depends on the UV cutoff $\Lambda$, \eqref{MLambda}, these mass limits imply lower limits on $\Lambda$.

\begin{figure}[ht]
    \includegraphics[width=0.6\linewidth]{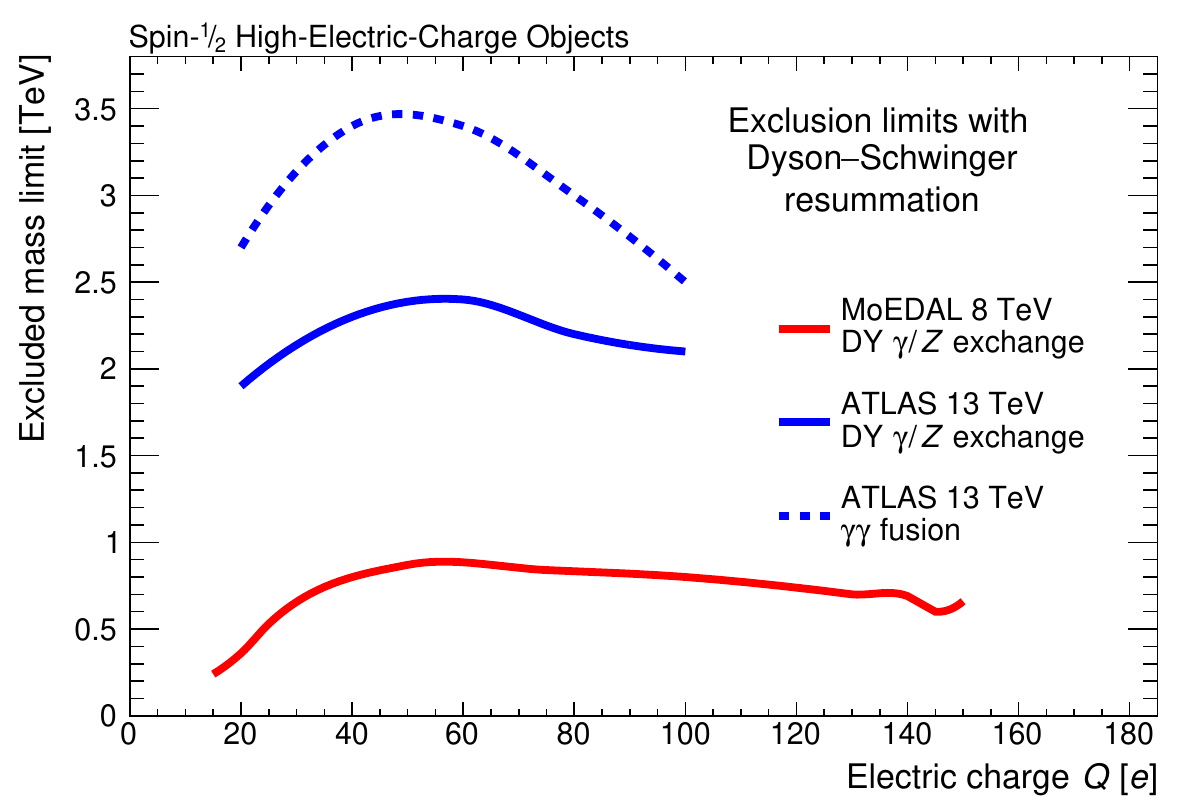}
    \caption{Re-interpreted excluded lower mass limits for spin-\hf HECOs from ATLAS~\cite{ATLAS:2015tyu,ATLAS:2023esy} and MoEDAL~\cite{MoEDAL:2021mpi}, considering the production cross sections with DS resummation obtained in this work.}
    \label{fig:hecos-resum}
\end{figure}

%%%%%%%%%%%%%%%%%%%%%%%%%%%%%%%%%%%%%%%%%%%%%%%%
%%%%%%%%%%%%%%%%%%%%%%%%%%%%%%%%%%%%%%%%%%%%%%%%
\section{Conclusions} \label{sec:concl} 

In this work, we have discussed resummation techniques that can be applied in the study of objects with high electric charge, the so-called HECOs. These hypothetical objects constitute the subject of active experimental searches at current colliders, as potential manifestations of physics beyond the SM. 
By employing specific Dyson-Schwinger techniques, we have arrived at an effective Lagrangian which can describe the dynamics of HECOs by means of appropriately dressed Drell-Yan and photon-fusion processes. This allows us to obtain theoretically reliable mass bounds and in general consistent data interpretation, from the pertinent experimentally acquired cross-section upper limits. As we have shown here, such techniques have significantly improved the associated mass bounds on HECOs obtained from collider searches, such as those performed by ATLAS and MoEDAL collaborations.

We implemented resummation effects on UFO models, compatible with Monte Carlo event generators such as \MAD, specifically focusing on the DY and PF production mechanisms involving spin-\hf  HECOs at the LHC. The first UFO model considers the contribution coming solely from photon exchange, while the second  includes also the $Z^0$ boson exchange, thus providing a more accurate representation of the physics involved. The application of these UFO models is not restricted to LHC only. It may be used to simulate events and calculate cross sections in future $e^+e^-$ Higgs factories, such as ILC, CLIC and CEPC, or other collider proposals such as the FCC-hh or a muon collider.

In order to include the above mentioned resummation effects, two new input parameters have been implemented: the multiplicity $n$ of the HECO charge $g = ne$ and the UV cutoff parameter $\Lambda$. These parameters influence the HECO mass and, consequently, affect the cross-section values. The improvement in production via the Drell-Yan process is approximately a factor of $\sim 2.1$, while the production via the photon-fusion process experiences an increase by a factor of $\sim 4.75$.
These results clearly demonstrate the positive impact of incorporating resummation effects on the predictive power of the UFO model. These outcomes have been used in this work to re-interpret such already made public results by collaborations, such as ATLAS and MoEDAL, that had searched for such objects by treating the production mechanisms at tree level only, to achieve more reliable and improved HECO mass bounds. 

Although in the current work we have discussed simplified HECO models, involving only non-chiral couplings of fermion HECOs to photons and $Z^0$, our analysis can be extended to more general models beyond the SM, involving also chiral coupling of the HECO to extra group factors that could characterise the underlying theory, as well as HECOs of other spins ---0 and 1--- that could exist in more general theories. We hope to come to such more general analysis in future publications.

%%%%%%%%%%%%%%%%%%%%%%%%%%%%%%%%%%%%%%%%%%%%%%%%
\section*{Acknowledgements}

The research of J.A.\ and N.E.M.\ is supported in part by the UK Science and Technology Facilities research Council (STFC) under the research grants ST/T000759/1 and ST/X000753/1, while that of V.A.M.\ and E.M.\ is supported by the Generalitat Valenciana via the Excellence Grant Prometeo CIPROM/2021/073 and by the Spanish MCIU / AEI / 10.13039/501100011033 and the European Union / FEDER via the grant PID2021-122134NB-C21. V.A.M.\ acknowledges support by the Spanish MCIU via the mobility grant PRX22/00633. E.M.\ acknowledges support by CSIC via the mobility grant iMOVE23097.

\FloatBarrier

\appendix 

%%%%%%%%%%%%%%%%%%%%%%%%%%%%%%%%%%%%%%%%%%%%%%%%
%%%%%%%%%%%%%%%%%%%%%%%%%%%%%%%%%%%%%%%%%%%%%%%%
\section{Feynman rules}\label{feynman_appendix}

In this Appendix, we provide details on the construction of the developed \MAD UFO models. Concerning the HECO interaction with photons only, the Feynman rules at tree level, used for the validation procedure presented in Appendix~\ref{validation_appendix}, are extracted from the Lagrangian in Eq.~\eqref{qedlag}. For the implementation of the resummation effects into \textsc{FeynRules}, we follow the relations introduced in Section~\ref{sec:hecorel}. Finally we derive the matrix amplitudes to calculate the final cross section for both DY and PF starting from the Feynman rules shown in Table~\ref{tab:feynmanrulesphotononly}. 
\begin{table}[ht]
  \caption{Feynman rules at tree level (second column) and after resummation (third column) for the case in which the spin-\hf HECO interacts with $\gamma$ only.}
  \label{tab:feynmanrulesphotononly}
  \centering
  \begin{tabular}{ c c c c c }
    \hline\hline
    && \multicolumn{3}{c}{\textbf{Feynman rules}}  \\
    \cline{3-5}
&& Tree level & & With resummation \\
    \hline
    \textbf{Vertices} & & & \\      
$u-\gamma-\Bar{u}$ 
    &&
    $-i Q_u e \gamma^\mu$ & & $-i Q_u e \gamma^\mu$\\
    $ \h-\gamma-\hb$ 
    &&
    $-i n e \gamma^\mu$ & & $-i n e \mathcal Z^* \gamma^\mu$ \\
    \hline
    \textbf{Propagators} & & & \\
    $\gamma$  \text{(Feynman gauge)}
    && $\frac{-i \eta_{\mu\nu}}{q^2}$ & & $\frac{-i}{q^2}\left(\eta_{\mu\nu}+ \omega^\star\,\frac{q_\mu q_\nu}{q^2}\right)$\\
    \h
    && $i \frac{\slashed p+M }{p^2-M^2}$ & & $i\frac{\slashed p+\mathcal M(\Lambda)}{p^2-\mathcal M(\Lambda)^2}$ \\
    \hline\hline
  \end{tabular}
\end{table}

The quantity $e$ is the elementary charge,  $Q_u e$ represents the electric charge of the up quark, which is $+2/3e$, $n$ denotes the multiplicity of the HECO charge, $q$ is the momentum transfer and $M$ represents the HECO mass. In addition, $\eta_{\mu\nu}$ denotes the metric tensor, $\gamma^\mu$ is the Dirac gamma matrix and $\theta_W$ stands for the weak mixing angle.  The quantities $\mathcal Z^*, \omega^*$ and $ \mathcal{M}(\Lambda)$ are defined in Eqs.~\eqref{ZomegaMbis},\eqref{omstasrdef} and \eqref{MLambda}, respectively.

We calculate the amplitudes for both DY and PF processes at tree level (Eqs.~\eqref{amplitude:dyp},\eqref{amplitude:pft},\eqref{amplitude:pfu}) and with resummation effects included (Eqs.~\eqref{amplitude:dypresum},\eqref{amplitude:pftresum},\eqref{amplitude:pfuresum}). The amplitude for the process $u \Bar{u} \to  \h \hb$ with $\gamma$-only exchange in the $s$-channel, at tree level, was obtained from the Feynman rules listed in the second column of Table~\ref{tab:feynmanrulesphotononly} and it can be written as
\begin{equation}
  -i  \mathcal{M}^\text{tree}_{\text{DY-}\gamma} = (-i\, e\, Q_u) (-i\, n\, e ) \bar{u}_{\h} \gamma^\mu v_{\hb} \frac{-i \eta_{\mu\nu}}{q^2}  \bar{v}_{\bar{u}} \gamma^\nu u_{u} \,, \label{amplitude:dyp}
\end{equation}
where $\bar{v}_{\bar{u}}$ expresses to the conjugate spinor for the initial anti-up-quark, while $u_{u}$ stands for the spinor associated with the initial up quark. Similarly, $\bar{u}_{\h}$ represents the conjugate spinor for the final HECO particle and $v_{\hb}$ denotes the spinor of the final anti-HECO particle. 

Let us now consider the HECO pair production via PF. The total matrix amplitude for the process is the sum of a $t$- and a $u$-channel process $\mathcal{M}_\text{PF}^\text{tree}= \mathcal{M}^\text{tree}_t+\mathcal{M}^\text{tree}_u$. The $t$-channel contribution to the amplitude for the PF process at tree level is given by
\begin{equation}
-i \mathcal{M}^\text{tree}_t = (-i \,n \,e)^2 \epsilon_{ \lambda_1 \mu}  \bar{u}_{\h}\gamma^{\mu} \frac{i (\slashed p_1 -\slashed p_3-M)}{(p_1-p_3)^2-M^2} \gamma^{\nu} v_{\hb} \epsilon_{ \lambda_2 \nu} \,,\label{amplitude:pft}
\end{equation}
where $\epsilon_{1,2}$ are the polarisation vectors of photons with momenta $p_1$ and $p_2$, and $\lambda_{1,2}$ represent the polarisation states, while $p_3$ and $p_4$ are the final-particles momenta.
The $u$-channel contribution to the amplitude is
\begin{equation}
-i \mathcal{M}^\text{tree}_u =  (-i\, n\, e)^2  \epsilon_{ \lambda_2 \mu}  \bar{u}_{\h} \gamma^{\mu} \frac{i (\slashed p_1 -\slashed p_4-M)}{(p_1-p_4)^2-M^2} \gamma^{\nu} v_{\hb} \epsilon_{ \lambda_1 \nu}\,.
\label{amplitude:pfu}
\end{equation}

By including the resummation effects, the amplitude for $u \Bar{u} \to  \h\hb$ with $\gamma$-only, is obtained from the Feynman rules in the third column of Table~\ref{tab:feynmanrulesphotononly} :
\begin{equation}
  -i  \mathcal{M}^\text{resum}_{\text{DY-}\gamma} = (-i\, e\, Q_u) (-i\, n \,\mathcal Z^*e ) \bar{u}_{\h} \gamma^\mu v_{\hb}  \frac{-i}{q^2}\left(\eta_{\mu\nu}+ \omega^\star \, \frac{q_\mu q_\nu}{q^2}\right)  \bar{v}_{\bar{u}} \gamma^\nu u_{u}. \label{amplitude:dypresum}
\end{equation}

For the PF process, the total matrix amplitude  is again the sum of a $t$- and a $u$-channel process:   $\mathcal{M}^{\text{resum}}_{\text{PF}}= \mathcal{M}^{\text{resum}}_t+\mathcal{M}^{\text{resum}}_u$. 
The $t$-channel contribution  is given by:
\begin{equation}
-i \mathcal{M}^{\text{resum}}_t = (-i\, n\, e \, \mathcal Z^*)^2  \epsilon_{ \lambda_1 \mu}  \bar{u}_{\h} \gamma^{\mu} \frac{i (\slashed p_1 -\slashed p_3-\mathcal{M}(\Lambda))}{(p_1-p_3)^2-\mathcal{M}(\Lambda)^2} \gamma^{\nu} v_{\hb} \epsilon_{ \lambda_2 \nu},
 \label{amplitude:pftresum}   
\end{equation}
while the $u$-channel contribution is
\begin{equation}
-i \mathcal{M}^{\text{resum}}_u =  (-i\, n \, e\, \mathcal Z^*)^2 \epsilon_{ \lambda_2 \mu} \bar{u}_{\h} \gamma^{\mu} \frac{i (\slashed p_1 -\slashed p_4-\mathcal{M}(\Lambda))}{(p_1-p_4)^2-\mathcal{M}(\Lambda)^2} \gamma^{\nu} v_{\hb} \epsilon_{ \lambda_1 \nu}. \label{amplitude:pfuresum}
\end{equation}

Let us now introduce the interaction of the HECO with the $Z^0$ boson. The Feynman rules at tree level are derived from Eq.~\eqref{nonchiralHECO}, while we  incorporated the resummation effects as discussed in Section~\ref{sec:zb}. We obtained the matrix amplitude used to compute the final cross section for DY with $Z^0$ exchange by utilising the Feynman rules provided in the Table~\ref{tab:feynmanrulesz}. 
\begin{table}[ht]
\caption{Feynman rules at tree level (second column) and after resummation (third column) when interaction with the $Z^0$~boson is included. }
\label{tab:feynmanrulesz}
  \centering
  \begin{tabular}{ c c c c c }
    \hline\hline
& & \multicolumn{3}{c}{\textbf{Feynman rules}}  \\
\cline{3-5}
& & Tree level & & With resummation \\
    \hline
    \textbf{Vertices} & & & & \\
    $u-\gamma-\Bar{u}$ 
    & & 
    $-i\, Q_u\, e\, \gamma^\mu$ & & $-i\, Q_u\, e\, \gamma^\mu$ \\
    $ \h-\gamma-\hb$ 
    & &
    $-i\, n\, e\, \gamma^\mu$ & & $-i\, n\, e\, \hat{\mathcal Z}^* \gamma^\mu$ \\
    $u-Z^0-\Bar{u}$  & & 
    $-i  \frac{e}{2 \sin \theta_W \cos \theta_W} \gamma^\mu \left(\frac{1}{2} - 2 Q_u \sin^2{\theta_W}- \frac{1}{2}\gamma^5\right)$ & & $-i  \frac{e}{2 \sin \theta_W \cos \theta_W} \gamma^\mu \left(\frac{1}{2} - 2 Q_u \sin^2{\theta_W}- \frac{1}{2}\gamma^5\right)$  \\
    $\h-Z^0-\hb$ 
   & &
    $-i  \frac{e}{2 \sin \theta_W \cos \theta_W} \gamma^\mu \left(\frac{1}{2}\zeta - 2 |n| \sin^2{\theta_W}-\zeta\gamma^5\right)$ & & $-i  \frac{e \hat{\mathcal Z}^*}{2 \sin \theta_W \cos \theta_W} \gamma^\mu \left(\frac{1}{2}\zeta - 2 |n| \sin^2{\theta_W}-\zeta\gamma^5\right)$ \\
   \hline
    \textbf{Propagators} & & & & \\
    $\gamma$
    \text{(Feynman gauge)}
    & & $\frac{-i \eta_{\mu\nu}}{q^2}$ & & $\frac{-i}{q^2}\left(\eta_{\mu\nu}+ \hat \omega^\star\,\frac{q_\mu q_\nu}{q^2}\right)$\\
    \h
    & & $i \frac{\slashed p+M }{p^2-M^2}$ & & $i\frac{\slashed p+\hat{\mathcal M}(\Lambda)}{p^2-(\hat{\mathcal M}(\Lambda))^2}$ \\
    $Z^0 $ 
   & & $\frac{-i }{q^2-M_Z^2} \left(\eta_{\mu\nu}-\frac{q_{\mu}q_{\nu}}{M_Z^2}\right)$ & & $\frac{-i }{q^2-M_Z^2} \left(\eta_{\mu\nu}- \frac{q_{\mu}q_{\nu}}{M_Z^2}\right)$ \\
   \hline\hline
  \end{tabular}
\end{table}

The mass of the $Z^0$ boson is denoted by $M_Z$, while $\hat{\mathcal{M}}(\Lambda)$ is the HECO running mass, which is introduced in Eq.~\eqref{mheco} together with the parameters $\hat{\mathcal Z}^*$ and $\hat{\omega}^*$.  The third component of the isospin value $I_3$ is $+1/2$ for the up-quark and 0 for the spin-\hf HECO. The $\zeta$ factor in the expression for the $\h-Z^0-\hb$ vertex is related to the type of HECO interactions. For vector interactions, which we consider here, $\zeta=0$, while it takes non-vanishing values for chiral interactions, e.g.\ $\zeta=1$ in case of left-handed HECOs.

In the case of $Z^0$ exchange in the HECO pair production via DY, according to the Feynman rules in Table~\ref{tab:feynmanrulesz}, the amplitude at tree level for the $Z^0$ contribution is 
\begin{equation}
\begin{aligned}
-i \mathcal{M}^\text{tree}_{\text{DY-}Z^0} = \left[ \bar{u}_{\h}  \frac{-ie}{2 \sin \theta_W \cos \theta_W} \gamma^\mu (\zeta - 2 |n|  \sin^2{\theta_W} - \zeta \gamma^5) v_{\hb} \right] 
\frac{-i }{q^2-M_Z^2} \left(\eta_{\mu\nu}-\frac{q_{\mu}q_{\nu}}{M_Z^2}\right) \\
\times \left[ \bar{v}_{\bar u} \frac{-ie}{2 \sin \theta_W \cos \theta_W} \gamma^\nu \left( \frac{1}{2} - \frac{4}{3}  \sin^2{\theta_W} - \frac{1}{2}\gamma^5 \right) u_u \right]. 
\end{aligned}
\label{amplitude:dyz}
\end{equation}
Considering the resummation effects in the third column of  Table~\ref{tab:feynmanrulesz}, one obtains the following amplitude:
\begin{equation}
\begin{aligned}
-i \mathcal{M}^\text{resum}_{\text{DY-}Z^0} =  \left[ \bar{u}_{\h}  \frac{-ie \mathcal Z^*}{2 \sin \theta_W \cos \theta_W} \gamma^\mu (\zeta - 2 |n|  \sin^2{\theta_W} - \zeta \gamma^5) v_{\hb} \right] 
\frac{-i }{q^2-M_Z^2} \left(\eta_{\mu\nu}-\frac{q_{\mu}q_{\nu}}{M_Z^2}\right) \\
\times \left[ \bar{v}_{\bar u} \frac{-ie}{2 \sin \theta_W \cos \theta_W} \gamma^\nu \left( \frac{1}{2} - \frac{4}{3}  \sin^2{\theta_W} - \frac{1}{2}\gamma^5 \right) u_u \right]. 
\end{aligned}
\label{amplitude:dyzresum}
\end{equation}

In the case of resummation, the following replacements have been applied to Eq.~\eqref{amplitude:dypresum}: $\mathcal Z^*\to\hat{\mathcal Z}^*$, $\omega^*\to\hat{\omega}^*$ and $\mathcal{M}(\Lambda) \to \hat{\mathcal{M}}(\Lambda)$. The final amplitude for DY HECO production with  $\gamma$ and $Z^0$ interactions is the sum $\mathcal{M}_\text{DY}=\mathcal{M}_{\text{DY-}\gamma}+\mathcal{M}_{\text{DY-}Z^0}$, considered for each case: at tree level and with resummation. 

%%%%%%%%%%%%%%%%%%%%%%%%%%%%%%%%%%%%%%%%%%%%%%%%
%%%%%%%%%%%%%%%%%%%%%%%%%%%%%%%%%%%%%%%%%%%%%%%%
\section{UFO model validation}\label{validation_appendix}

To validate the UFO models for HECOs described in Section~\ref{sec:feyn}, cross-section values obtained from the analytical calculations using the \Math~13.0.1 package \Feyncalc~9.3.1~\cite{Feyncalc} are compared to those calculated by \madamc. 
This validation procedure aims at ensuring the consistency and accuracy of the implemented UFO models for HECOs, providing confidence in its predictive capabilities.

The validation process with \Math started by considering the scattering amplitudes for the DY process, separately for photon-only and for $Z^0$-boson-included interactions. Once the amplitude expressions were defined, we proceeded to compute their complex conjugates, the squared amplitudes and eventually the probability amplitudes.  In order to facilitate the calculation, the Mandelstam variables were introduced, in order to characterise the energy and momentum transfers within the scattering process. From the squared amplitudes we obtained the differential cross sections with respect to the solid angle. Further, the UFO validation involved the integration over the phase space, accounting for the momenta and energies of the particles involved in the interaction. The integrated expression became our benchmark variable allowing comparisons between various parameters, level (tree or resummation), production mechanisms and calculation methods.

Before embarking in validating the developed UFO models that include the resummation effects, we validated the tree-level estimates through the total HECO-pair-production cross section. We confronted the results obtained from \MAD with the corresponding analytical outcome of the theoretical model obtained using \Math. For validation purposes, the initial partons ---up-quarks or photons--- are considered to collide directly, hence no PDF was included in the simulation, i.e.\ the \texttt{no-PDF} option is set in \MAD. 

The comparison between the \MAD and \Math results related to DY  collision  processes of $u \bar u$ quarks into \h \hb  ~at tree level is shown in Table~\ref{tabDY_treelevel} for both cases; $\gamma$  exchange only and $\gamma/Z^0$ exchange. Similar validation results for HECO-pair production via PF at tree level are presented in Table~\ref{tab:photonfusiontree}. In these tables, the UFO/theory ratio is also provided, which turns out to be close to the unity at the per-mille level, confirming that the simulation code reproduces accurately the  theory at tree level. 
\begin{table}[ht]
  \caption{Cross-section of spin-\hf HECO production of mass $M=1~\tev$ via the DY $u \Bar{u} \to \h \hb$ process at $\sqrt{s}=13~\tev$ obtained with \MAD and with theoretical calculations at tree level. The simulation/theory ratio is also listed.}
  \label{tabDY_treelevel}
  \centering
\begin{tabular}{ c c c c c c c c c }
\hline\hline
\multicolumn{9}{ c }{DY $u \bar{u}\to \h\hb$ tree level @ $\sqrt{s}=13~\tev$, $M= 1~\tev$ }\\
\hline
\multirow{2}{*}{$Q\,(e)$}&&\multicolumn{3}{c}{$\gamma$-only exchange}&&\multicolumn{3}{c}{$\gamma$/$Z^0$ exchange}\\ \cline{3-5}\cline{7-9}
 && $\sigma_\MAD$ (pb) & $\sigma_\Math$  (pb)&  UFO/Theory & &$\sigma_\MAD$ (pb) & $\sigma_\Math$  (pb)&  UFO/Theory\\
\hline
20 & \phantom{99} & 0.0349 & 0.0348 & 1.003 & \phantom{99} & 0.0315  & 0.0314  & 1.003 \\
40 && 0.1395   & 0.1390 & 1.007 &&  0.1262    & 0.1257 & 1.004\\
60 &&  0.3141 & 0.3127 & 1.005 && 0.2845 & 0.2828 & 1.006\\
80 &&  0.5579  & 0.5560 & 1.004 && 0.5056 &0.5028 & 1.005\\
100 && 0.8748 & 0.8687 & 1.004 && 0.7903& 0.7857& 1.006\\
120 && 1.2589 & 1.2509 & 1.005 && 1.1371 & 1.1313 & 1.005 \\
140 && 1.7142 & 1.7026& 1.006 && 1.5482 & 1.5399& 1.005\\
160 && 2.2332 & 2.2238 & 1.007 &&  2.0229 & 2.0113 & 1.006\\
180 && 2.8277& 2.8145 & 1.005 && 2.5620 &2.5455& 1.006\\
200 && 3.4926 & 3.4747 & 1.003 &&  3.1536& 3.1426 & 1.003\\
\hline\hline
\end{tabular}
\end{table}

\begin{table}[ht]
  \caption{Cross-section of spin-\hf HECO production of mass $M=1~\tev$ via the PF $\gamma\gamma \to \h\hb$  process at $\sqrt{s}=13~\tev$ obtained with \MAD and with theoretical calculations at tree level. The simulation/theory ratio is also listed.}
  \label{tab:photonfusiontree}
  \centering
  \begin{tabular}{ >{\centering}p{2.1cm}>{\centering}p{3.3cm}>{\centering}p{3.3cm}>{\centering\arraybackslash}p{3.3cm} }
    \hline\hline
    \multicolumn{4}{c}{ PF tree level $\gamma\gamma \to \h\hb$  @ $\sqrt{s}=13~\tev$, $M = 1~\tev$ }  \\
    \hline
    $Q~(e)$ & $\sigma_{\MAD}$ (pb) & $\sigma_{\Math}$ (pb) & UFO/Theory \\
    \hline
20 & 7.438$\times10^3$ & 7.398$\times10^3$ & 1.005 \\
40 & 7.732$\times10^4$ & 7.692$\times10^4$ & 1.004 \\
60 & 3.539$\times10^5$ & 3.528$\times10^5$ & 1.003 \\
80 & 1.082$\times10^6$ & 1.076$\times10^6$ & 1.006 \\
100 & 2.596$\times10^6$ & 2.580$\times10^9$ & 1.006 \\
120 & 5.327$\times10^6$ & 5.300$\times10^6$ & 1.005 \\
140 & 9.814$\times10^6$ & 9.761$\times10^6$ & 1.005 \\
160 & 1.668$\times10^7$ & 1.659$\times10^7$ & 1.005 \\
180 & 2.663$\times10^7$ & 2.651$\times10^7$ & 1.004 \\
200 & 4.052$\times10^7$ & 4.033$\times10^7$ & 1.005 \\
    \hline\hline
  \end{tabular}
\end{table}

We now proceed to validate the actual UFO models, which include DS resummation effects, following exactly the same procedure as for the tree level. The cross-section values including resummation effects for the DY process are reported in Table~\ref{tabDY}. It is reminded that the DY with exchange of photons only utilises the respective UFO model, while when $Z^0$-boson exchange is also considered, the alternative UFO model is deployed. In a similar fashion, the validation results for the PF process are summarised in Table~\ref{tabPF}. For both processes and UFO models, good agreement with theory is observed. Therefore, the validity of the implemented resummation effects in the UFO models is confirmed and these tools can be reliably utilised for simulating high-energy particle collisions with the implemented resummation effects in the UFO models.

\begin{table*}[ht]
\caption{Cross-section of spin-\hf HECO production via the DY $u \Bar{u} \to \h \hb$ process at $\sqrt{s}=13~\tev$ obtained with \MAD by importing the relevant UFO models and with theoretical calculations. Resummation with $\Lambda= 2~\tev$ is assumed. The simulation/theory ratio is also listed.}
\label{tabDY}
\centering
\begin{tabular}{ c c c c c c c c c }
\hline\hline
\multicolumn{9}{ c }{DY with resummation $u \bar{u}\to \h\hb$ @ $\sqrt{s}=13~\tev$, $\Lambda= 2~\tev$}\\
\hline
\multirow{2}{*}{$Q\,(e)$}&&\multicolumn{3}{c}{$\gamma$-only exchange}&&\multicolumn{3}{c}{$\gamma$/$Z^0$ exchange}\\ \cline{3-5}\cline{7-9}
 && $\sigma_\MAD$ (pb) & $\sigma_\Math$  (pb)&  UFO/Theory & &$\sigma_\MAD$ (pb) & $\sigma_\Math$  (pb)&  UFO/Theory\\
\hline
20 & \phantom{99} & 0.0762 & 0.0758 & 1.005 & \phantom{99} & 0.0659 & 0.0655  & 1.006 \\
40 && 0.3048   & 0.3027 & 1.007 &&  0.2632 & 0.2616 & 1.006\\
60 &&  0.6837 & 0.6807 & 1.004 && 0.5919 &0.5886 &  1.005\\
80 &&  1.2151 & 1.2097 & 1.004 && 1.0508 & 1.0462 & 1.004\\
100 && 1.8976 & 1.8898 & 1.004 && 1.6460 &  1.6345 & 1.007\\
120 && 2.7372 & 2.7210 & 1.006 && 2.3682 & 2.3537 & 1.006 \\
140 && 3.7261  & 3.7032 & 1.006 && 3.2251 & 3.2035 & 1.006\\
160 && 4.8712  & 4.8366 & 1.007 &&  4.2083 & 4.1843 & 1.005\\
180 && 6.1548 & 6.1210 & 1.006 && 5.3224 & 5.2955 & 1.005\\
200 && 7.5927  & 7.5568 & 1.004 &&  6.5758 & 6.5379 & 1.006\\
\hline\hline
\end{tabular}
\end{table*}

\begin{table}[ht]
\caption{Cross-section of spin-\hf HECO production via the PF $\gamma\gamma \to \h\hb$  process at $\sqrt{s}=13~\tev$ obtained with \MAD by importing the relevant UFO models and with theoretical calculations. Resummation with $\Lambda= 2~\tev$ is assumed. The simulation/theory ratio is also listed.}
\label{tabPF}
\centering
\begin{tabular}{ >{\centering}p{1.1cm}>{\centering}p{3.3cm}>{\centering}p{3.3cm}>{\centering\arraybackslash}p{2.3cm}  }
\hline\hline
\multicolumn{4}{ c }{ PF with resummation $\gamma\gamma \to \h\hb$  @ $\sqrt{s}=13~\tev$, $\Lambda = 2~\tev$} \\
\hline
$Q~(e)$ & $\sigma_\MAD$ (pb) & $\sigma_\Math$  (pb)&  UFO/Theory\\
\hline
20 & 7.438$\times10^3$ & 7.398$\times10^3$ & 1.005 \\
40 & 7.732$\times10^4$ & 7.692$\times10^4$ & 1.004 \\
60 & 3.539$\times10^5$ & 3.528$\times10^5$ & 1.003 \\
80 & 1.082$\times10^6$ & 1.076$\times10^6$ & 1.006 \\
100 & 2.596$\times10^6$ & 2.580$\times10^9$ & 1.006 \\
120 & 5.327$\times10^6$ & 5.300$\times10^6$ & 1.005 \\
140 & 9.814$\times10^6$ & 9.761$\times10^6$ & 1.005 \\
160 & 1.668$\times10^7$ & 1.659$\times10^7$ & 1.005 \\
180 & 2.663$\times10^7$ & 2.651$\times10^7$ & 1.004 \\
200 & 4.052$\times10^7$ & 4.033$\times10^7$ & 1.005 \\
\hline\hline
\end{tabular} 
\end{table}

\FloatBarrier

%%%%%%%%%%%%%%%%%%%%%%%%%%%%%%%%%%%%%%%%%%%%%%%%
\bibliographystyle{apsrev4-1}
\bibliography{heco}

\begin{thebibliography}{44}%
\makeatletter
\providecommand \@ifxundefined [1]{%
 \@ifx{#1\undefined}
}%
\providecommand \@ifnum [1]{%
 \ifnum #1\expandafter \@firstoftwo
 \else \expandafter \@secondoftwo
 \fi
}%
\providecommand \@ifx [1]{%
 \ifx #1\expandafter \@firstoftwo
 \else \expandafter \@secondoftwo
 \fi
}%
\providecommand \natexlab [1]{#1}%
\providecommand \enquote  [1]{``#1''}%
\providecommand \bibnamefont  [1]{#1}%
\providecommand \bibfnamefont [1]{#1}%
\providecommand \citenamefont [1]{#1}%
\providecommand \href@noop [0]{\@secondoftwo}%
\providecommand \href [0]{\begingroup \@sanitize@url \@href}%
\providecommand \@href[1]{\@@startlink{#1}\@@href}%
\providecommand \@@href[1]{\endgroup#1\@@endlink}%
\providecommand \@sanitize@url [0]{\catcode `\\12\catcode `\$12\catcode
  `\&12\catcode `\#12\catcode `\^12\catcode `\_12\catcode `\%12\relax}%
\providecommand \@@startlink[1]{}%
\providecommand \@@endlink[0]{}%
\providecommand \url  [0]{\begingroup\@sanitize@url \@url }%
\providecommand \@url [1]{\endgroup\@href {#1}{\urlprefix }}%
\providecommand \urlprefix  [0]{URL }%
\providecommand \Eprint [0]{\href }%
\providecommand \doibase [0]{http://dx.doi.org/}%
\providecommand \selectlanguage [0]{\@gobble}%
\providecommand \bibinfo  [0]{\@secondoftwo}%
\providecommand \bibfield  [0]{\@secondoftwo}%
\providecommand \translation [1]{[#1]}%
\providecommand \BibitemOpen [0]{}%
\providecommand \bibitemStop [0]{}%
\providecommand \bibitemNoStop [0]{.\EOS\space}%
\providecommand \EOS [0]{\spacefactor3000\relax}%
\providecommand \BibitemShut  [1]{\csname bibitem#1\endcsname}%
\let\auto@bib@innerbib\@empty
%</preamble>
\bibitem [{\citenamefont {Acharya}\ \emph {et~al.}(2014)\citenamefont {Acharya}
  \emph {et~al.}}]{dcmp}%
  \BibitemOpen
  \bibfield  {author} {\bibinfo {author} {\bibfnamefont {B.}~\bibnamefont
  {Acharya}} \emph {et~al.} (\bibinfo {collaboration} {MoEDAL}),\ }\href
  {\doibase 10.1142/S0217751X14300506} {\bibfield  {journal} {\bibinfo
  {journal} {Int. J. Mod. Phys. A}\ }\textbf {\bibinfo {volume} {29}},\
  \bibinfo {pages} {1430050} (\bibinfo {year} {2014})},\ \Eprint
  {http://arxiv.org/abs/1405.7662} {arXiv:1405.7662 [hep-ph]} \BibitemShut
  {NoStop}%
\bibitem [{\citenamefont {Schwinger}(1969)}]{dyons}%
  \BibitemOpen
  \bibfield  {author} {\bibinfo {author} {\bibfnamefont {J.~S.}\ \bibnamefont
  {Schwinger}},\ }\href {\doibase 10.1126/science.165.3895.757} {\bibfield
  {journal} {\bibinfo  {journal} {Science}\ }\textbf {\bibinfo {volume}
  {165}},\ \bibinfo {pages} {757} (\bibinfo {year} {1969})}\BibitemShut
  {NoStop}%
\bibitem [{\citenamefont {Hirsch}\ \emph {et~al.}(2021)\citenamefont {Hirsch},
  \citenamefont {Mase\l{}ek},\ and\ \citenamefont {Sakurai}}]{scalarneutrino}%
  \BibitemOpen
  \bibfield  {author} {\bibinfo {author} {\bibfnamefont {M.}~\bibnamefont
  {Hirsch}}, \bibinfo {author} {\bibfnamefont {R.}~\bibnamefont {Mase\l{}ek}},
  \ and\ \bibinfo {author} {\bibfnamefont {K.}~\bibnamefont {Sakurai}},\ }\href
  {\doibase 10.1140/epjc/s10052-021-09507-9} {\bibfield  {journal} {\bibinfo
  {journal} {Eur. Phys. J. C}\ }\textbf {\bibinfo {volume} {81}},\ \bibinfo
  {pages} {697} (\bibinfo {year} {2021})},\ \bibinfo {note} {[Erratum:
  Eur.Phys.J.C 82, 774 (2022)]},\ \Eprint {http://arxiv.org/abs/2103.05644}
  {arXiv:2103.05644 [hep-ph]} \BibitemShut {NoStop}%
\bibitem [{\citenamefont {Coleman}(1985)}]{coleman}%
  \BibitemOpen
  \bibfield  {author} {\bibinfo {author} {\bibfnamefont {S.~R.}\ \bibnamefont
  {Coleman}},\ }\href {\doibase 10.1016/0550-3213(85)90286-X} {\bibfield
  {journal} {\bibinfo  {journal} {Nucl. Phys. B}\ }\textbf {\bibinfo {volume}
  {262}},\ \bibinfo {pages} {263} (\bibinfo {year} {1985})},\ \bibinfo {note}
  {[Addendum: Nucl.Phys.B 269, 744 (1986)]}\BibitemShut {NoStop}%
\bibitem [{\citenamefont {Kusenko}\ and\ \citenamefont
  {Shaposhnikov}(1998)}]{kusenko}%
  \BibitemOpen
  \bibfield  {author} {\bibinfo {author} {\bibfnamefont {A.}~\bibnamefont
  {Kusenko}}\ and\ \bibinfo {author} {\bibfnamefont {M.~E.}\ \bibnamefont
  {Shaposhnikov}},\ }\href {\doibase 10.1016/S0370-2693(97)01375-0} {\bibfield
  {journal} {\bibinfo  {journal} {Phys. Lett. B}\ }\textbf {\bibinfo {volume}
  {418}},\ \bibinfo {pages} {46} (\bibinfo {year} {1998})},\ \Eprint
  {http://arxiv.org/abs/hep-ph/9709492} {arXiv:hep-ph/9709492} \BibitemShut
  {NoStop}%
\bibitem [{\citenamefont {Holdom}\ \emph {et~al.}(2018)\citenamefont {Holdom},
  \citenamefont {Ren},\ and\ \citenamefont {Zhang}}]{udaggr}%
  \BibitemOpen
  \bibfield  {author} {\bibinfo {author} {\bibfnamefont {B.}~\bibnamefont
  {Holdom}}, \bibinfo {author} {\bibfnamefont {J.}~\bibnamefont {Ren}}, \ and\
  \bibinfo {author} {\bibfnamefont {C.}~\bibnamefont {Zhang}},\ }\href
  {\doibase 10.1103/PhysRevLett.120.222001} {\bibfield  {journal} {\bibinfo
  {journal} {Phys. Rev. Lett.}\ }\textbf {\bibinfo {volume} {120}},\ \bibinfo
  {pages} {222001} (\bibinfo {year} {2018})},\ \Eprint
  {http://arxiv.org/abs/1707.06610} {arXiv:1707.06610 [hep-ph]} \BibitemShut
  {NoStop}%
\bibitem [{\citenamefont {Farhi}\ and\ \citenamefont {Jaffe}(1984)}]{smatter}%
  \BibitemOpen
  \bibfield  {author} {\bibinfo {author} {\bibfnamefont {E.}~\bibnamefont
  {Farhi}}\ and\ \bibinfo {author} {\bibfnamefont {R.~L.}\ \bibnamefont
  {Jaffe}},\ }\href {\doibase 10.1103/PhysRevD.30.2379} {\bibfield  {journal}
  {\bibinfo  {journal} {Phys. Rev. D}\ }\textbf {\bibinfo {volume} {30}},\
  \bibinfo {pages} {2379} (\bibinfo {year} {1984})}\BibitemShut {NoStop}%
\bibitem [{\citenamefont {Alberghi}\ \emph {et~al.}(2013)\citenamefont
  {Alberghi}, \citenamefont {Bellagamba}, \citenamefont {Calmet}, \citenamefont
  {Casadio},\ and\ \citenamefont {Micu}}]{cbhrem}%
  \BibitemOpen
  \bibfield  {author} {\bibinfo {author} {\bibfnamefont {G.~L.}\ \bibnamefont
  {Alberghi}}, \bibinfo {author} {\bibfnamefont {L.}~\bibnamefont
  {Bellagamba}}, \bibinfo {author} {\bibfnamefont {X.}~\bibnamefont {Calmet}},
  \bibinfo {author} {\bibfnamefont {R.}~\bibnamefont {Casadio}}, \ and\
  \bibinfo {author} {\bibfnamefont {O.}~\bibnamefont {Micu}},\ }\href {\doibase
  10.1140/epjc/s10052-013-2448-0} {\bibfield  {journal} {\bibinfo  {journal}
  {Eur. Phys. J. C}\ }\textbf {\bibinfo {volume} {73}},\ \bibinfo {pages}
  {2448} (\bibinfo {year} {2013})},\ \Eprint {http://arxiv.org/abs/1303.3150}
  {arXiv:1303.3150 [hep-ph]} \BibitemShut {NoStop}%
\bibitem [{\citenamefont {Koch}\ \emph {et~al.}(2007)\citenamefont {Koch},
  \citenamefont {Bleicher},\ and\ \citenamefont {Stoecker}}]{bhrem}%
  \BibitemOpen
  \bibfield  {author} {\bibinfo {author} {\bibfnamefont {B.}~\bibnamefont
  {Koch}}, \bibinfo {author} {\bibfnamefont {M.}~\bibnamefont {Bleicher}}, \
  and\ \bibinfo {author} {\bibfnamefont {H.}~\bibnamefont {Stoecker}},\ }\href
  {\doibase 10.1088/0954-3899/34/8/S44} {\bibfield  {journal} {\bibinfo
  {journal} {J. Phys. G}\ }\textbf {\bibinfo {volume} {34}},\ \bibinfo {pages}
  {S535} (\bibinfo {year} {2007})},\ \Eprint
  {http://arxiv.org/abs/hep-ph/0702187} {arXiv:hep-ph/0702187} \BibitemShut
  {NoStop}%
\bibitem [{\citenamefont {Song}\ and\ \citenamefont {Taylor}(2022)}]{Song}%
  \BibitemOpen
  \bibfield  {author} {\bibinfo {author} {\bibfnamefont {W.~Y.}\ \bibnamefont
  {Song}}\ and\ \bibinfo {author} {\bibfnamefont {W.}~\bibnamefont {Taylor}},\
  }\href {\doibase 10.1088/1361-6471/ac3dce} {\bibfield  {journal} {\bibinfo
  {journal} {J. Phys. G}\ }\textbf {\bibinfo {volume} {49}},\ \bibinfo {pages}
  {045002} (\bibinfo {year} {2022})},\ \Eprint
  {http://arxiv.org/abs/2107.10789} {arXiv:2107.10789 [hep-ph]} \BibitemShut
  {NoStop}%
\bibitem [{\citenamefont {Aad}\ \emph {et~al.}(2016)\citenamefont {Aad} \emph
  {et~al.}}]{ATLAS:2015tyu}%
  \BibitemOpen
  \bibfield  {author} {\bibinfo {author} {\bibfnamefont {G.}~\bibnamefont
  {Aad}} \emph {et~al.} (\bibinfo {collaboration} {ATLAS}),\ }\href {\doibase
  10.1103/PhysRevD.93.052009} {\bibfield  {journal} {\bibinfo  {journal} {Phys.
  Rev. D}\ }\textbf {\bibinfo {volume} {93}},\ \bibinfo {pages} {052009}
  (\bibinfo {year} {2016})},\ \Eprint {http://arxiv.org/abs/1509.08059}
  {arXiv:1509.08059 [hep-ex]} \BibitemShut {NoStop}%
\bibitem [{\citenamefont {Aad}\ \emph {et~al.}(2020)\citenamefont {Aad} \emph
  {et~al.}}]{ATLAS:2019wkg}%
  \BibitemOpen
  \bibfield  {author} {\bibinfo {author} {\bibfnamefont {G.}~\bibnamefont
  {Aad}} \emph {et~al.} (\bibinfo {collaboration} {ATLAS}),\ }\href {\doibase
  10.1103/PhysRevLett.124.031802} {\bibfield  {journal} {\bibinfo  {journal}
  {Phys. Rev. Lett.}\ }\textbf {\bibinfo {volume} {124}},\ \bibinfo {pages}
  {031802} (\bibinfo {year} {2020})},\ \Eprint
  {http://arxiv.org/abs/1905.10130} {arXiv:1905.10130 [hep-ex]} \BibitemShut
  {NoStop}%
\bibitem [{\citenamefont {Aad}\ \emph {et~al.}(2023{\natexlab{a}})\citenamefont
  {Aad} \emph {et~al.}}]{ATLAS:2023esy}%
  \BibitemOpen
  \bibfield  {author} {\bibinfo {author} {\bibfnamefont {G.}~\bibnamefont
  {Aad}} \emph {et~al.} (\bibinfo {collaboration} {ATLAS}),\ }\href {\doibase
  10.1007/JHEP11(2023)112} {\bibfield  {journal} {\bibinfo  {journal} {JHEP}\
  }\textbf {\bibinfo {volume} {11}},\ \bibinfo {pages} {112} (\bibinfo {year}
  {2023}{\natexlab{a}})},\ \Eprint {http://arxiv.org/abs/2308.04835}
  {arXiv:2308.04835 [hep-ex]} \BibitemShut {NoStop}%
\bibitem [{\citenamefont {Acharya}\ \emph {et~al.}(2022)\citenamefont {Acharya}
  \emph {et~al.}}]{MoEDAL:2021mpi}%
  \BibitemOpen
  \bibfield  {author} {\bibinfo {author} {\bibfnamefont {B.}~\bibnamefont
  {Acharya}} \emph {et~al.} (\bibinfo {collaboration} {MoEDAL}),\ }\href
  {\doibase 10.1140/epjc/s10052-022-10608-2} {\bibfield  {journal} {\bibinfo
  {journal} {Eur. Phys. J. C}\ }\textbf {\bibinfo {volume} {82}},\ \bibinfo
  {pages} {694} (\bibinfo {year} {2022})},\ \Eprint
  {http://arxiv.org/abs/2112.05806} {arXiv:2112.05806 [hep-ex]} \BibitemShut
  {NoStop}%
\bibitem [{\citenamefont {Aad}\ \emph {et~al.}(2015)\citenamefont {Aad} \emph
  {et~al.}}]{ATLAS:2015hau}%
  \BibitemOpen
  \bibfield  {author} {\bibinfo {author} {\bibfnamefont {G.}~\bibnamefont
  {Aad}} \emph {et~al.} (\bibinfo {collaboration} {ATLAS}),\ }\href {\doibase
  10.1140/epjc/s10052-015-3534-2} {\bibfield  {journal} {\bibinfo  {journal}
  {Eur. Phys. J. C}\ }\textbf {\bibinfo {volume} {75}},\ \bibinfo {pages} {362}
  (\bibinfo {year} {2015})},\ \Eprint {http://arxiv.org/abs/1504.04188}
  {arXiv:1504.04188 [hep-ex]} \BibitemShut {NoStop}%
\bibitem [{\citenamefont {Aaboud}\ \emph {et~al.}(2019)\citenamefont {Aaboud}
  \emph {et~al.}}]{ATLAS:2018imb}%
  \BibitemOpen
  \bibfield  {author} {\bibinfo {author} {\bibfnamefont {M.}~\bibnamefont
  {Aaboud}} \emph {et~al.} (\bibinfo {collaboration} {ATLAS}),\ }\href
  {\doibase 10.1103/PhysRevD.99.052003} {\bibfield  {journal} {\bibinfo
  {journal} {Phys. Rev. D}\ }\textbf {\bibinfo {volume} {99}},\ \bibinfo
  {pages} {052003} (\bibinfo {year} {2019})},\ \Eprint
  {http://arxiv.org/abs/1812.03673} {arXiv:1812.03673 [hep-ex]} \BibitemShut
  {NoStop}%
\bibitem [{\citenamefont {Aad}\ \emph {et~al.}(2023{\natexlab{b}})\citenamefont
  {Aad} \emph {et~al.}}]{ATLAS:2023zxo}%
  \BibitemOpen
  \bibfield  {author} {\bibinfo {author} {\bibfnamefont {G.}~\bibnamefont
  {Aad}} \emph {et~al.} (\bibinfo {collaboration} {ATLAS}),\ }\href {\doibase
  10.1016/j.physletb.2023.138316} {\bibfield  {journal} {\bibinfo  {journal}
  {Phys. Lett. B}\ }\textbf {\bibinfo {volume} {847}},\ \bibinfo {pages}
  {138316} (\bibinfo {year} {2023}{\natexlab{b}})},\ \Eprint
  {http://arxiv.org/abs/2303.13613} {arXiv:2303.13613 [hep-ex]} \BibitemShut
  {NoStop}%
\bibitem [{\citenamefont {Chatrchyan}\ \emph {et~al.}(2013)\citenamefont
  {Chatrchyan} \emph {et~al.}}]{CMS:2013czn}%
  \BibitemOpen
  \bibfield  {author} {\bibinfo {author} {\bibfnamefont {S.}~\bibnamefont
  {Chatrchyan}} \emph {et~al.} (\bibinfo {collaboration} {CMS}),\ }\href
  {\doibase 10.1007/JHEP07(2013)122} {\bibfield  {journal} {\bibinfo  {journal}
  {JHEP}\ }\textbf {\bibinfo {volume} {07}},\ \bibinfo {pages} {122} (\bibinfo
  {year} {2013})},\ \bibinfo {note} {[Erratum: JHEP 11, 149 (2022)]},\ \Eprint
  {http://arxiv.org/abs/1305.0491} {arXiv:1305.0491 [hep-ex]} \BibitemShut
  {NoStop}%
\bibitem [{\citenamefont {Altakach}\ \emph {et~al.}(2022)\citenamefont
  {Altakach}, \citenamefont {Lamba}, \citenamefont {Mase\l{}ek}, \citenamefont
  {Mitsou},\ and\ \citenamefont {Sakurai}}]{Altakach:2022hgn}%
  \BibitemOpen
  \bibfield  {author} {\bibinfo {author} {\bibfnamefont {M.~M.}\ \bibnamefont
  {Altakach}}, \bibinfo {author} {\bibfnamefont {P.}~\bibnamefont {Lamba}},
  \bibinfo {author} {\bibfnamefont {R.}~\bibnamefont {Mase\l{}ek}}, \bibinfo
  {author} {\bibfnamefont {V.~A.}\ \bibnamefont {Mitsou}}, \ and\ \bibinfo
  {author} {\bibfnamefont {K.}~\bibnamefont {Sakurai}},\ }\href {\doibase
  10.1140/epjc/s10052-022-10805-z} {\bibfield  {journal} {\bibinfo  {journal}
  {Eur. Phys. J. C}\ }\textbf {\bibinfo {volume} {82}},\ \bibinfo {pages} {848}
  (\bibinfo {year} {2022})},\ \Eprint {http://arxiv.org/abs/2204.03667}
  {arXiv:2204.03667 [hep-ph]} \BibitemShut {NoStop}%
\bibitem [{\citenamefont {Mavromatos}\ and\ \citenamefont
  {Mitsou}(2020)}]{Mavromatos:2020gwk}%
  \BibitemOpen
  \bibfield  {author} {\bibinfo {author} {\bibfnamefont {N.~E.}\ \bibnamefont
  {Mavromatos}}\ and\ \bibinfo {author} {\bibfnamefont {V.~A.}\ \bibnamefont
  {Mitsou}},\ }\href {\doibase 10.1142/S0217751X20300124} {\bibfield  {journal}
  {\bibinfo  {journal} {Int. J. Mod. Phys. A}\ }\textbf {\bibinfo {volume}
  {35}},\ \bibinfo {pages} {2030012} (\bibinfo {year} {2020})},\ \Eprint
  {http://arxiv.org/abs/2005.05100} {arXiv:2005.05100 [hep-ph]} \BibitemShut
  {NoStop}%
\bibitem [{\citenamefont {Schwinger}\ \emph {et~al.}(1976)\citenamefont
  {Schwinger}, \citenamefont {Milton}, \citenamefont {Tsai}, \citenamefont
  {DeRaad},\ and\ \citenamefont {Clark}}]{milton1}%
  \BibitemOpen
  \bibfield  {author} {\bibinfo {author} {\bibfnamefont {J.~S.}\ \bibnamefont
  {Schwinger}}, \bibinfo {author} {\bibfnamefont {K.~A.}\ \bibnamefont
  {Milton}}, \bibinfo {author} {\bibfnamefont {W.-y.}\ \bibnamefont {Tsai}},
  \bibinfo {author} {\bibfnamefont {L.~L.}\ \bibnamefont {DeRaad},
  \bibfnamefont {Jr.}}, \ and\ \bibinfo {author} {\bibfnamefont {D.~C.}\
  \bibnamefont {Clark}},\ }\href {\doibase 10.1016/0003-4916(76)90020-8}
  {\bibfield  {journal} {\bibinfo  {journal} {Annals Phys.}\ }\textbf {\bibinfo
  {volume} {101}},\ \bibinfo {pages} {451} (\bibinfo {year}
  {1976})}\BibitemShut {NoStop}%
\bibitem [{\citenamefont {Milton}(2006)}]{milton2}%
  \BibitemOpen
  \bibfield  {author} {\bibinfo {author} {\bibfnamefont {K.~A.}\ \bibnamefont
  {Milton}},\ }\href {\doibase 10.1088/0034-4885/69/6/R02} {\bibfield
  {journal} {\bibinfo  {journal} {Rept. Prog. Phys.}\ }\textbf {\bibinfo
  {volume} {69}},\ \bibinfo {pages} {1637} (\bibinfo {year} {2006})},\ \Eprint
  {http://arxiv.org/abs/hep-ex/0602040} {arXiv:hep-ex/0602040} \BibitemShut
  {NoStop}%
\bibitem [{\citenamefont {Baines}\ \emph {et~al.}(2018)\citenamefont {Baines},
  \citenamefont {Mavromatos}, \citenamefont {Mitsou}, \citenamefont {Pinfold},\
  and\ \citenamefont {Santra}}]{baines}%
  \BibitemOpen
  \bibfield  {author} {\bibinfo {author} {\bibfnamefont {S.}~\bibnamefont
  {Baines}}, \bibinfo {author} {\bibfnamefont {N.~E.}\ \bibnamefont
  {Mavromatos}}, \bibinfo {author} {\bibfnamefont {V.~A.}\ \bibnamefont
  {Mitsou}}, \bibinfo {author} {\bibfnamefont {J.~L.}\ \bibnamefont {Pinfold}},
  \ and\ \bibinfo {author} {\bibfnamefont {A.}~\bibnamefont {Santra}},\ }\href
  {\doibase 10.1140/epjc/s10052-018-6440-6} {\bibfield  {journal} {\bibinfo
  {journal} {Eur. Phys. J. C}\ }\textbf {\bibinfo {volume} {78}},\ \bibinfo
  {pages} {966} (\bibinfo {year} {2018})},\ \bibinfo {note} {[Erratum:
  Eur.Phys.J.C 79, 166 (2019)]},\ \Eprint {http://arxiv.org/abs/1808.08942}
  {arXiv:1808.08942 [hep-ph]} \BibitemShut {NoStop}%
\bibitem [{\citenamefont {Gusynin}\ \emph {et~al.}(1999)\citenamefont
  {Gusynin}, \citenamefont {Miransky},\ and\ \citenamefont
  {Shovkovy}}]{Gusynin:1999pq}%
  \BibitemOpen
  \bibfield  {author} {\bibinfo {author} {\bibfnamefont {V.~P.}\ \bibnamefont
  {Gusynin}}, \bibinfo {author} {\bibfnamefont {V.~A.}\ \bibnamefont
  {Miransky}}, \ and\ \bibinfo {author} {\bibfnamefont {I.~A.}\ \bibnamefont
  {Shovkovy}},\ }\href {\doibase 10.1016/S0550-3213(99)00573-8} {\bibfield
  {journal} {\bibinfo  {journal} {Nucl. Phys. B}\ }\textbf {\bibinfo {volume}
  {563}},\ \bibinfo {pages} {361} (\bibinfo {year} {1999})},\ \Eprint
  {http://arxiv.org/abs/hep-ph/9908320} {arXiv:hep-ph/9908320} \BibitemShut
  {NoStop}%
\bibitem [{\citenamefont {Aguilar}\ \emph {et~al.}(2016)\citenamefont
  {Aguilar}, \citenamefont {Binosi},\ and\ \citenamefont
  {Papavassiliou}}]{Aguilar:2015bud}%
  \BibitemOpen
  \bibfield  {author} {\bibinfo {author} {\bibfnamefont {A.~C.}\ \bibnamefont
  {Aguilar}}, \bibinfo {author} {\bibfnamefont {D.}~\bibnamefont {Binosi}}, \
  and\ \bibinfo {author} {\bibfnamefont {J.}~\bibnamefont {Papavassiliou}},\
  }\href {\doibase 10.1007/s11467-015-0517-6} {\bibfield  {journal} {\bibinfo
  {journal} {Front. Phys. (Beijing)}\ }\textbf {\bibinfo {volume} {11}},\
  \bibinfo {pages} {111203} (\bibinfo {year} {2016})},\ \Eprint
  {http://arxiv.org/abs/1511.08361} {arXiv:1511.08361 [hep-ph]} \BibitemShut
  {NoStop}%
\bibitem [{\citenamefont {Alexandre}\ and\ \citenamefont
  {Mavromatos}(2019)}]{AM}%
  \BibitemOpen
  \bibfield  {author} {\bibinfo {author} {\bibfnamefont {J.}~\bibnamefont
  {Alexandre}}\ and\ \bibinfo {author} {\bibfnamefont {N.~E.}\ \bibnamefont
  {Mavromatos}},\ }\href {\doibase 10.1103/PhysRevD.100.096005} {\bibfield
  {journal} {\bibinfo  {journal} {Phys. Rev. D}\ }\textbf {\bibinfo {volume}
  {100}},\ \bibinfo {pages} {096005} (\bibinfo {year} {2019})},\ \Eprint
  {http://arxiv.org/abs/1906.08738} {arXiv:1906.08738 [hep-ph]} \BibitemShut
  {NoStop}%
\bibitem [{\citenamefont {Alwall}\ \emph {et~al.}(2011)\citenamefont {Alwall},
  \citenamefont {Herquet}, \citenamefont {Maltoni}, \citenamefont {Mattelaer},\
  and\ \citenamefont {Stelzer}}]{madgraph}%
  \BibitemOpen
  \bibfield  {author} {\bibinfo {author} {\bibfnamefont {J.}~\bibnamefont
  {Alwall}}, \bibinfo {author} {\bibfnamefont {M.}~\bibnamefont {Herquet}},
  \bibinfo {author} {\bibfnamefont {F.}~\bibnamefont {Maltoni}}, \bibinfo
  {author} {\bibfnamefont {O.}~\bibnamefont {Mattelaer}}, \ and\ \bibinfo
  {author} {\bibfnamefont {T.}~\bibnamefont {Stelzer}},\ }\href {\doibase
  10.1007/JHEP06(2011)128} {\bibfield  {journal} {\bibinfo  {journal} {JHEP}\
  }\textbf {\bibinfo {volume} {06}},\ \bibinfo {pages} {128} (\bibinfo {year}
  {2011})},\ \Eprint {http://arxiv.org/abs/1106.0522} {arXiv:1106.0522
  [hep-ph]} \BibitemShut {NoStop}%
\bibitem [{\citenamefont {Cornwall}\ and\ \citenamefont
  {Papavassiliou}(1989)}]{pinched}%
  \BibitemOpen
  \bibfield  {author} {\bibinfo {author} {\bibfnamefont {J.~M.}\ \bibnamefont
  {Cornwall}}\ and\ \bibinfo {author} {\bibfnamefont {J.}~\bibnamefont
  {Papavassiliou}},\ }\href {\doibase 10.1103/PhysRevD.40.3474} {\bibfield
  {journal} {\bibinfo  {journal} {Phys. Rev. D}\ }\textbf {\bibinfo {volume}
  {40}},\ \bibinfo {pages} {3474} (\bibinfo {year} {1989})}\BibitemShut
  {NoStop}%
\bibitem [{\citenamefont {Binosi}\ and\ \citenamefont
  {Papavassiliou}(2009)}]{pinched2}%
  \BibitemOpen
  \bibfield  {author} {\bibinfo {author} {\bibfnamefont {D.}~\bibnamefont
  {Binosi}}\ and\ \bibinfo {author} {\bibfnamefont {J.}~\bibnamefont
  {Papavassiliou}},\ }\href {\doibase 10.1016/j.physrep.2009.05.001} {\bibfield
   {journal} {\bibinfo  {journal} {Phys. Rept.}\ }\textbf {\bibinfo {volume}
  {479}},\ \bibinfo {pages} {1} (\bibinfo {year} {2009})},\ \Eprint
  {http://arxiv.org/abs/0909.2536} {arXiv:0909.2536 [hep-ph]} \BibitemShut
  {NoStop}%
\bibitem [{\citenamefont {Alexandre}\ \emph {et~al.}(2002)\citenamefont
  {Alexandre}, \citenamefont {Polonyi},\ and\ \citenamefont {Sailer}}]{APS}%
  \BibitemOpen
  \bibfield  {author} {\bibinfo {author} {\bibfnamefont {J.}~\bibnamefont
  {Alexandre}}, \bibinfo {author} {\bibfnamefont {J.}~\bibnamefont {Polonyi}},
  \ and\ \bibinfo {author} {\bibfnamefont {K.}~\bibnamefont {Sailer}},\ }\href
  {\doibase 10.1016/S0370-2693(02)01482-X} {\bibfield  {journal} {\bibinfo
  {journal} {Phys. Lett. B}\ }\textbf {\bibinfo {volume} {531}},\ \bibinfo
  {pages} {316} (\bibinfo {year} {2002})},\ \Eprint
  {http://arxiv.org/abs/hep-th/0111152} {arXiv:hep-th/0111152} \BibitemShut
  {NoStop}%
\bibitem [{\citenamefont {Gallagher}\ \emph {et~al.}(2020)\citenamefont
  {Gallagher}, \citenamefont {Groote},\ and\ \citenamefont
  {Naeem}}]{Gallagher:2020ajd}%
  \BibitemOpen
  \bibfield  {author} {\bibinfo {author} {\bibfnamefont {P.}~\bibnamefont
  {Gallagher}}, \bibinfo {author} {\bibfnamefont {S.}~\bibnamefont {Groote}}, \
  and\ \bibinfo {author} {\bibfnamefont {M.}~\bibnamefont {Naeem}},\ }\href
  {\doibase 10.3390/particles3030037} {\bibfield  {journal} {\bibinfo
  {journal} {Particles}\ }\textbf {\bibinfo {volume} {3}},\ \bibinfo {pages}
  {543} (\bibinfo {year} {2020})},\ \Eprint {http://arxiv.org/abs/2001.04106}
  {arXiv:2001.04106 [hep-ph]} \BibitemShut {NoStop}%
\bibitem [{\citenamefont {Fornal}\ \emph {et~al.}(2018)\citenamefont {Fornal},
  \citenamefont {Manohar},\ and\ \citenamefont {Waalewijn}}]{Fornal:2018znf}%
  \BibitemOpen
  \bibfield  {author} {\bibinfo {author} {\bibfnamefont {B.}~\bibnamefont
  {Fornal}}, \bibinfo {author} {\bibfnamefont {A.~V.}\ \bibnamefont {Manohar}},
  \ and\ \bibinfo {author} {\bibfnamefont {W.~J.}\ \bibnamefont {Waalewijn}},\
  }\href {\doibase 10.1007/JHEP05(2018)106} {\bibfield  {journal} {\bibinfo
  {journal} {JHEP}\ }\textbf {\bibinfo {volume} {05}},\ \bibinfo {pages} {106}
  (\bibinfo {year} {2018})},\ \Eprint {http://arxiv.org/abs/1803.06347}
  {arXiv:1803.06347 [hep-ph]} \BibitemShut {NoStop}%
\bibitem [{\citenamefont {Mangano}\ \emph {et~al.}(2016)\citenamefont {Mangano}
  \emph {et~al.}}]{Mangano:2016jyj}%
  \BibitemOpen
  \bibfield  {author} {\bibinfo {author} {\bibfnamefont {M.~L.}\ \bibnamefont
  {Mangano}} \emph {et~al.},\ }\href {\doibase 10.23731/CYRM-2017-003.1} {\
  (\bibinfo {year} {2016}),\ 10.23731/CYRM-2017-003.1},\ \Eprint
  {http://arxiv.org/abs/1607.01831} {arXiv:1607.01831 [hep-ph]} \BibitemShut
  {NoStop}%
\bibitem [{\citenamefont {Godunov}\ \emph {et~al.}(2023)\citenamefont
  {Godunov}, \citenamefont {Karkaryan}, \citenamefont {Novikov}, \citenamefont
  {Rozanov}, \citenamefont {Vysotsky},\ and\ \citenamefont
  {Zhemchugov}}]{Godunov:2023myj}%
  \BibitemOpen
  \bibfield  {author} {\bibinfo {author} {\bibfnamefont {S.~I.}\ \bibnamefont
  {Godunov}}, \bibinfo {author} {\bibfnamefont {E.~K.}\ \bibnamefont
  {Karkaryan}}, \bibinfo {author} {\bibfnamefont {V.~A.}\ \bibnamefont
  {Novikov}}, \bibinfo {author} {\bibfnamefont {A.~N.}\ \bibnamefont
  {Rozanov}}, \bibinfo {author} {\bibfnamefont {M.~I.}\ \bibnamefont
  {Vysotsky}}, \ and\ \bibinfo {author} {\bibfnamefont {E.~V.}\ \bibnamefont
  {Zhemchugov}},\ }\href {\doibase 10.1103/PhysRevD.108.093006} {\bibfield
  {journal} {\bibinfo  {journal} {Phys. Rev. D}\ }\textbf {\bibinfo {volume}
  {108}},\ \bibinfo {pages} {093006} (\bibinfo {year} {2023})},\ \Eprint
  {http://arxiv.org/abs/2308.01169} {arXiv:2308.01169 [hep-ph]} \BibitemShut
  {NoStop}%
\bibitem [{\citenamefont {Alloul}\ \emph {et~al.}(2014)\citenamefont {Alloul},
  \citenamefont {Christensen}, \citenamefont {Degrande}, \citenamefont {Duhr},\
  and\ \citenamefont {Fuks}}]{Alloul:2013bka}%
  \BibitemOpen
  \bibfield  {author} {\bibinfo {author} {\bibfnamefont {A.}~\bibnamefont
  {Alloul}}, \bibinfo {author} {\bibfnamefont {N.~D.}\ \bibnamefont
  {Christensen}}, \bibinfo {author} {\bibfnamefont {C.}~\bibnamefont
  {Degrande}}, \bibinfo {author} {\bibfnamefont {C.}~\bibnamefont {Duhr}}, \
  and\ \bibinfo {author} {\bibfnamefont {B.}~\bibnamefont {Fuks}},\ }\href
  {\doibase 10.1016/j.cpc.2014.04.012} {\bibfield  {journal} {\bibinfo
  {journal} {Comput. Phys. Commun.}\ }\textbf {\bibinfo {volume} {185}},\
  \bibinfo {pages} {2250} (\bibinfo {year} {2014})},\ \Eprint
  {http://arxiv.org/abs/1310.1921} {arXiv:1310.1921 [hep-ph]} \BibitemShut
  {NoStop}%
\bibitem [{\citenamefont {Degrande}\ \emph {et~al.}(2012)\citenamefont
  {Degrande}, \citenamefont {Duhr}, \citenamefont {Fuks}, \citenamefont
  {Grellscheid}, \citenamefont {Mattelaer},\ and\ \citenamefont
  {Reiter}}]{ufo}%
  \BibitemOpen
  \bibfield  {author} {\bibinfo {author} {\bibfnamefont {C.}~\bibnamefont
  {Degrande}}, \bibinfo {author} {\bibfnamefont {C.}~\bibnamefont {Duhr}},
  \bibinfo {author} {\bibfnamefont {B.}~\bibnamefont {Fuks}}, \bibinfo {author}
  {\bibfnamefont {D.}~\bibnamefont {Grellscheid}}, \bibinfo {author}
  {\bibfnamefont {O.}~\bibnamefont {Mattelaer}}, \ and\ \bibinfo {author}
  {\bibfnamefont {T.}~\bibnamefont {Reiter}},\ }\href {\doibase
  10.1016/j.cpc.2012.01.022} {\bibfield  {journal} {\bibinfo  {journal}
  {Comput. Phys. Commun.}\ }\textbf {\bibinfo {volume} {183}},\ \bibinfo
  {pages} {1201} (\bibinfo {year} {2012})},\ \Eprint
  {http://arxiv.org/abs/1108.2040} {arXiv:1108.2040 [hep-ph]} \BibitemShut
  {NoStop}%
\bibitem [{\citenamefont {Ball}\ \emph {et~al.}(2013)\citenamefont {Ball} \emph
  {et~al.}}]{NPDF}%
  \BibitemOpen
  \bibfield  {author} {\bibinfo {author} {\bibfnamefont {R.~D.}\ \bibnamefont
  {Ball}} \emph {et~al.},\ }\href {\doibase 10.1016/j.nuclphysb.2012.10.003}
  {\bibfield  {journal} {\bibinfo  {journal} {Nucl. Phys. B}\ }\textbf
  {\bibinfo {volume} {867}},\ \bibinfo {pages} {244} (\bibinfo {year}
  {2013})},\ \Eprint {http://arxiv.org/abs/1207.1303} {arXiv:1207.1303
  [hep-ph]} \BibitemShut {NoStop}%
\bibitem [{\citenamefont {Manohar}\ \emph {et~al.}(2017)\citenamefont
  {Manohar}, \citenamefont {Nason}, \citenamefont {Salam},\ and\ \citenamefont
  {Zanderighi}}]{luxqed}%
  \BibitemOpen
  \bibfield  {author} {\bibinfo {author} {\bibfnamefont {A.~V.}\ \bibnamefont
  {Manohar}}, \bibinfo {author} {\bibfnamefont {P.}~\bibnamefont {Nason}},
  \bibinfo {author} {\bibfnamefont {G.~P.}\ \bibnamefont {Salam}}, \ and\
  \bibinfo {author} {\bibfnamefont {G.}~\bibnamefont {Zanderighi}},\ }\href
  {\doibase 10.1007/JHEP12(2017)046} {\bibfield  {journal} {\bibinfo  {journal}
  {JHEP}\ }\textbf {\bibinfo {volume} {12}},\ \bibinfo {pages} {046} (\bibinfo
  {year} {2017})},\ \Eprint {http://arxiv.org/abs/1708.01256} {arXiv:1708.01256
  [hep-ph]} \BibitemShut {NoStop}%
\bibitem [{\citenamefont {Aad}\ \emph {et~al.}(2008)\citenamefont {Aad} \emph
  {et~al.}}]{ATLAS:2008xda}%
  \BibitemOpen
  \bibfield  {author} {\bibinfo {author} {\bibfnamefont {G.}~\bibnamefont
  {Aad}} \emph {et~al.} (\bibinfo {collaboration} {ATLAS}),\ }\href {\doibase
  10.1088/1748-0221/3/08/S08003} {\bibfield  {journal} {\bibinfo  {journal}
  {JINST}\ }\textbf {\bibinfo {volume} {3}},\ \bibinfo {pages} {S08003}
  (\bibinfo {year} {2008})}\BibitemShut {NoStop}%
\bibitem [{\citenamefont {Pinfold}\ \emph {et~al.}(2009)\citenamefont {Pinfold}
  \emph {et~al.}}]{MoEDAL:2009jwa}%
  \BibitemOpen
  \bibfield  {author} {\bibinfo {author} {\bibfnamefont {J.}~\bibnamefont
  {Pinfold}} \emph {et~al.} (\bibinfo {collaboration} {MoEDAL}),\ }\href@noop
  {} {\enquote {\bibinfo {title} {{Technical Design Report of the MoEDAL
  Experiment}},}\ } (\bibinfo {year} {2009}),\ \bibinfo {note}
  {{CERN-LHCC-2009-006, MoEDAL-TDR-001}}\BibitemShut {NoStop}%
\bibitem [{\citenamefont {Mitsou}(2003)}]{Mitsou:2003rp}%
  \BibitemOpen
  \bibfield  {author} {\bibinfo {author} {\bibfnamefont {V.~A.}\ \bibnamefont
  {Mitsou}} (\bibinfo {collaboration} {ATLAS TRT}),\ }in\ \href {\doibase
  10.1142/9789812702708_0073} {\emph {\bibinfo {booktitle} {{8th International
  Conference on Advanced Technology and Particle Physics (ICATPP 2003):
  Astroparticle, Particle, Space Physics, Detectors and Medical Physics
  Applications}}}}\ (\bibinfo {year} {2003})\ pp.\ \bibinfo {pages}
  {497--501},\ \Eprint {http://arxiv.org/abs/hep-ex/0311058}
  {arXiv:hep-ex/0311058} \BibitemShut {NoStop}%
\bibitem [{\citenamefont {Aad}\ \emph {et~al.}(2010)\citenamefont {Aad} \emph
  {et~al.}}]{ATLAS:2010blk}%
  \BibitemOpen
  \bibfield  {author} {\bibinfo {author} {\bibfnamefont {G.}~\bibnamefont
  {Aad}} \emph {et~al.} (\bibinfo {collaboration} {ATLAS}),\ }\href {\doibase
  10.1140/epjc/s10052-010-1354-y} {\bibfield  {journal} {\bibinfo  {journal}
  {Eur. Phys. J. C}\ }\textbf {\bibinfo {volume} {70}},\ \bibinfo {pages} {723}
  (\bibinfo {year} {2010})},\ \Eprint {http://arxiv.org/abs/0912.2642}
  {arXiv:0912.2642 [physics.ins-det]} \BibitemShut {NoStop}%
\bibitem [{\citenamefont {Acharya}\ \emph {et~al.}(2023)\citenamefont {Acharya}
  \emph {et~al.}}]{MoEDAL:2023ost}%
  \BibitemOpen
  \bibfield  {author} {\bibinfo {author} {\bibfnamefont {B.}~\bibnamefont
  {Acharya}} \emph {et~al.} (\bibinfo {collaboration} {MoEDAL}),\ }\href@noop
  {} {\  (\bibinfo {year} {2023})},\ \Eprint {http://arxiv.org/abs/2311.06509}
  {arXiv:2311.06509 [hep-ex]} \BibitemShut {NoStop}%
\bibitem [{\citenamefont {Shtabovenko}\ \emph {et~al.}(2020)\citenamefont
  {Shtabovenko}, \citenamefont {Mertig},\ and\ \citenamefont
  {Orellana}}]{Feyncalc}%
  \BibitemOpen
  \bibfield  {author} {\bibinfo {author} {\bibfnamefont {V.}~\bibnamefont
  {Shtabovenko}}, \bibinfo {author} {\bibfnamefont {R.}~\bibnamefont {Mertig}},
  \ and\ \bibinfo {author} {\bibfnamefont {F.}~\bibnamefont {Orellana}},\
  }\href {\doibase 10.1016/j.cpc.2020.107478} {\bibfield  {journal} {\bibinfo
  {journal} {Comput. Phys. Commun.}\ }\textbf {\bibinfo {volume} {256}},\
  \bibinfo {pages} {107478} (\bibinfo {year} {2020})},\ \Eprint
  {http://arxiv.org/abs/2001.04407} {arXiv:2001.04407 [hep-ph]} \BibitemShut
  {NoStop}%
\end{thebibliography}%

\end{document}
